\documentstyle[]{article}
\font\cero=cmss10 scaled 1728
\font\uno=cmssbx10 scaled 1200
\begin{document}
\small{
\begin{flushleft}
{\cero Adjoint operators, gauge invariant perturbations, and covariant 
symplectic structure for black holes in string theory}  \\[3em]
\end{flushleft}
{\sf R. Cartas-Fuentevilla} \\
{\it Instituto de F\'{\i}sica, Universidad Aut\'onoma de Puebla,
Apartado postal J-48 72570 Puebla Pue., M\'exico} \\ [4em]

Expressions for the general and complete perturbations in terms of Debye 
potentials of static charged black holes in string theory, valid for 
curvature below the Planck scale, are derived starting from a decoupled 
set of equations and using Wald's method of adjoint operators. Our results  
cover both extremal and nonextremal black holes and are valid for 
arbitrary values of the dilaton coupling parameter. The decoupled set is 
obtained using the Newman-Penrose formulation of the 
Einstein-Maxwell-dilaton theory and involves naturally field quantities 
invariant under both ordinary gauge transformations of the electromagnetic
potential perturbations and infinitesimal rotations of the perturbed 
tetrad. Furthermore, using the recent pointed out relationship between 
adjoint operators and conserved currents, a local continuity law for the 
field perturbations in terms of the potentials is also obtained. It is 
shown that such continuity equation implies the existence of conserved 
quantities and of a covariant symplectic structure on the phase space. 
Future extensions of the present results are discussed.

\noindent PACS numbers: 04.20.Jb, 04.40.Nr\\
\noindent Keywords: Adjoint operators, perturbations, symplectic
structure, black holes, string theory, Debye potentials.\\
\noindent Running title: Adjoint operators.....\\

\begin{center}
{\uno I. INTRODUCTION}
\end{center}
\vspace{1em}

At present, the theories of extended objects such as membranes and strings
represent the more viable candidates for the quantum theory of gravity. 
Particularly, there have been many efforts studying black holes in string
theory from different points of view, with the main task of elucidating on 
the problem of quantum gravity embedded in them, since such objects appear 
to play a crucial role in the subject. However, because of the many 
technical and conceptual difficulties in treating the full theory, the 
low-energy limit of string theory has been developed as a more pragmatic 
approach. This low-energy physics emerges as an effective action obtained 
from the lowest order in the world-sheet and string loop expansion, where 
the usual Einstein-Hilbert gravity is supplemented by gauge fields, scalar 
fields such as the axion and the dilaton, which couple in a nontrivial way 
to the other matter fields \cite{1}. As it is well known, the presence of 
the dilaton changes drastically the dynamical properties of the systems, 
and new features arise in this theory due to the nontrivial coupling of 
this field. In particular, dilaton black holes have shown to have novel 
thermodynamics properties \cite{2,3}, and to behave like elementary 
particles in the sense that the excitation spectrum has an energy gap 
\cite{4,5,6}. Besides, it has been explored the viewpoint that quantum
black holes are massive excitations of extended objects and also 
correspond, in this sense, to elementary particles ([7], and references 
cited therein).

On the experimental context, recent investigations attempt to explore a
possible experimental evidence of string theory. Since string theory
predicts particularly the existence of the dilaton scalar field, the new
generation of detectors of gravity waves are sensitive in the presence of
a possible scalar component of such waves. Specifically, a scalar
component of gravity radiation should excite the monopole mode of new
resonant-mass detectors of spherical shape \cite{8}, and should give a
especific correlation between an interferometer and the monopole mode of a
resonant sphere \cite{9}. Furthermore, the spherical resonant-mass
detectors \cite{10}, or an array of interferometers \cite{11} are able, in
principle, to determine the spin content of the incoming gravitational
waves possibly coupled with their scalar components. In this same context,
black holes should be the more typical and possible astrophysical source
of gravity waves.

In all issues discussed above, the first-order perturbation analysis plays
a fundamental role. Perturbation theory revels important physical
information of the system under study. As we shall see, the adjoint
operators approach will cover, in an unified way, various aspects of the
same problem (in this case, the perturbation analysis of string black
holes), which traditionally have been treated separately. In the remainder
of this Introduction, we discuss such aspects, pointing out our aims and
successes in the present work, and we make a review of previous works in
which the present approach has been employed.

In the scheme of the perturbation theory, the black holes (and other 
spacetimes) have been studied from different approaches. The traditional 
approaches consist to try of solving the original set of equations for the 
field perturbations directly. This approach has several disadvantages and 
difficulties that can be overcame by means of an alternative and more 
convenient approach based on the concept of the adjoint of a differential 
operator (Wald's method). The reach and differences of this approach with
respect to the usual ones have been already discussed widely in previous 
works (see for example \cite{12}, and references therein). In fact, in the 
cases where string fields are involved, the approach has been applied 
successfully in the setting of the Einstein-Maxwell-dilaton-axion (EMDA) 
theory, which contains the low-energy limit of string theory as a 
particular case \cite{12,13}. Additionally, as we shall see, with the
connection recently established between adjoint operators and conserved 
currents, Wald's method becomes the more convenient and powerful approach
for facing the study of perturbations.

At a more general context, the study of conservation laws in field 
theories involving gravity, becomes particularly interesting because of 
the lack of conserved currents representing the conservation of energy 
and momentum. Additionally, in the construction of a {\it covariant 
symplectic structure} on the phase space of classical systems, a bilinear
product on first-order deformations of classical solutions on such phase
space is required. In both cases, the problem is to find a local
expression physically meaningful and coming from some continuity equation. 
As we shall see, the present adjoint operators scheme allows us to
establish a local continuity law with the features described above, from 
which conserved quantities and a covariant symplectic structure (in terms 
of Debye potentials) are derived.

It is important to emphasize, at this point, the significance of a
covariant symplectic structure in field theory. As well known, Feynman 
path integral and canonical quantization are the fundamental approaches 
in quantum field theories. If quantization is carried out by means of
path integral, the resultant theory has no necessarily the standard 
structure in terms of quantum mechanical states and operators. In fact, in 
string field theory, the existence of such a structure is not obvious 
\cite{14}. However, Feynman path integral has the great virtue of 
preserving manifestly the Poincar\'e invariance. As opposed to path 
integral, the canonical formalism, with a suitable definition of Poisson 
brackets, leads to Hamiltonian mechanics of the standard form, which yields 
a quantum theory of the conventional type (replacing Poisson brackets with
conmutators). Although this formalism usually is considered that does not 
preserve the Poincar\'e invariance, Witten \cite {14}, Crncovi\'c and
Witten \cite {15}, and Suckerman \cite{16} have achieved to describe
Poisson brackets in terms of a symplectic structure on the classical phase 
space in a covariant way. In such description, the classical phase space
is defined as {\it the space of solutions of the classical equations of 
motion}; such definition is manifestly covariant. The construction of a 
covariantly conserved two-form $J^{\mu}$ on such phase space yields a 
symplectic structure $\omega$ defined as $\omega \equiv \int_{\Sigma} 
J^{\mu} d \Sigma_{\mu}$, being $\Sigma$ an initial value hypersurface, 
independent of the choice of $\Sigma$ and, in particular, Poincar\'e 
invariant. Additionally, in terms of symplectic structure $\omega$, the 
fact that Poisson brackets satisfy the Jacoby identity, is equivalent that  
$\omega$ to be a closed two-form on the phase space, which holds if 
$J^{\mu}$ itself is closed. With this properties, $J^{\mu}$ is known as 
{\it the symplectic current}. One of our goals in the present paper is to 
establish a local continuity equation that permits to identify, in a 
straightforward way, {\it a symplectic current} for the solution 
considered.

In this manner, the purpose of the present work is to perform an analysis
of the first-order perturbations of the dilatonic charged black holes
employing Wald's method. Previously, it has been demonstrated the
self-adjointness of the operator governing the field perturbations in the  
EMDA theory \cite{12,13}, remaining only the finding of the corresponding 
decoupled set of equations in the case where the background space-time 
corresponds to the solution considered, in order to establish our results.

For this purpose, the outline of this paper is as follows. Section II is
dedicated to establish the general relationship between adjoint operators 
and conserved currents, and the extensions of the original Wald's method; 
some issues on the notation are also discussed in this Section. The 
relevant information on the background solution is given in Sec.\ III. In 
Section IV, a decoupled set of equations for metric, vector potential, and 
dilaton perturbations is obtained from the original equations for the field
perturbations, which are given in Appendix A using the Newman-Penrose
formulation. Employing the results of Section IV, the equations for the
Debye potentials, and the expressions for the metric, vector potential,
and dilaton perturbations in terms of those, are found in Sec.\ 5.1. In
Sec.\ 5.2, our fundamental continuity equation is established and a
symplectic structure is derived in Sec.\ 5.3. Some additional comments on  
the role that the Debye potentials play in the present approach, are given
in Sec.\ 5.3. The separation of variables for the equations for the Debye  
potentials, and for the continuity equation is performed in Sec.\ VI,
such that two conserved quantities are obtained. We conclude this Section
with certain differential identities and we comment briefly on their
meaning. Appendix B is useful in this section. Finally, we finish with
some concluding remarks and future extensions of the present results. \\[2em]

\noindent {\uno II. ADJOINT OPERATORS}
\vspace{1em}

\noindent{\uno 2.1 New branch of adjoint operators: local continuity laws}
\vspace{.5cm}

In Refs. \cite{17} it has been shown that there exists a conserved
current associated with any system of homogeneous linear partial
differential equations that can be written in terms of a self-adjoint
operator. This result is limited for a self-adjoint system, for which the
corresponding conserved current depends on a pair of solutions admitted by
such a system. However, as we shall see below, there exists a more general
possibility that extends for systems of equations that are not
self-adjoint necessarily. The demonstration is very easy (see also 
\cite{18}):

In accordance with Wald's definition \cite{19}, if ${\cal E}$ corresponds
to a linear partial differential operator which maps $m$-index tensor
fields into $n$-index tensor fields, then, the adjoint operator of $\cal 
E$, denoted by ${\cal E}^{\dag}$, is that linear partial differential 
operator mapping $n$-index tensor fields into $m$-index tensor fields such 
that
\begin{equation}
     g^{\rho\sigma\ldots} [{\cal E} (f_{\mu\nu\ldots})]_{\rho\sigma\ldots}
     - [{\cal E}^{\dag} (g^{\rho\sigma\ldots})]^{\mu\nu\ldots}
     f_{\mu\nu\ldots} = \nabla_{\mu} J^{\mu},
\end{equation}
where $J^{\mu}$ is some vector field depending on the fields $f$ and $g$.
From Eq.\ (1) we can see that this definition automatically guarantees
that, if the field $f$ is a solution of the linear system ${\cal E} (f) = 
0$ and $g$ a solution of the adjoint system ${\cal E}^{\dag} (g) = 0$,
then $J^{\mu}$ is a covariantly conserved current. This fact means that 
for any homogeneous equation system, one can always construct a conserved 
current taking into account the adjoint system. This general result 
contains the self-adjoint case as a particular one. 

In the present work, $f$ and $g$ will be associated  with the first-order
variations of the backgrounds fields. Such field variations will
correspond, on the phase space, to one-forms \cite{15}. In this manner,
the left-hand side of Eq.\ (1) can be understood as a wedge product on 
such phase space: $g \wedge {\cal E}(f) - {\cal E}^{\dag}(g) \wedge f =
\nabla_{\mu} J^{\mu}$, and something similar for the bilinear form
$J^{\mu}$ in its dependence on the fields $f$ and $g$ (the operators
${\cal E}$, ${\cal E}^{\dag}$, and $\nabla_{\mu}$ will depend only on
the background fields).

It is worth pointing out some issues on the notation. The first-order
field variations appearing in Refs. \cite{12,13} are denoted by a
superscript B. On the other hand, the field variations coincide, in 
according to Witten's interpretation \cite{15}, with an
infinite-dimensional generalization of the usual exterior derivative, 
which is traditionally represented by the symbol $\delta$. However, in 
Refs. \cite{12,13} and present work, the Newman-Penrose formalism is used, 
in which the symbol $\delta$ is employed for denoting one of the 
directional derivatives defined by the null tetrad. In this manner, for 
avoiding confusion, we will maintain the symbol $\delta$ as usual in the 
Newman-Penrose notation, and the superscript B for the first-order field 
variations (the exterior derivative of background fields). In the present 
article, the exterior derivative will not be performed explicitly, and it 
will be sufficient for our purposes to understand any quantity with the 
superscript B as a one-form on the phase space. Quantities without such a 
superscript will correspond to background fields, which mean zero-forms on 
the phase space. With these previous considerations, formulae and
notation of Refs. \cite{12,13} will be used throughout this paper; the 
concepts and definitions on differential forms, exterior derivatives, etc, 
come from Ref.\ \cite{15}. \\[2em]

\noindent{\uno 2.2 Traditional branch of adjoint operators: decoupled
equations and potentials}
\vspace{.5cm}

For completeness, we outline the original idea for introducing the
definition (1) in Ref.\ \cite{19}: reduction of systems of linear partial
differential equations to equations for scalar potentials (called Debye
potentials), which determine a complete solution of the original system.

If we have the linear system ${\cal E}(f) = 0$, and there exist linear
operators such that
\[
   {\cal S} {\cal E} = {\cal O} {\cal T}, \nonumber
\]
identically, then the field ${\cal S}^{\dag} (\psi)$ satisfies the
equation
\[
   {\cal E}^{\dag} ({\cal S}^{\dag}(\psi)) = 0, \nonumber
\]
provided that the scalar field $\psi$ satisfies
\[
   {\cal O}^{\dag} (\psi) = 0. \nonumber
\]

In particular, if ${\cal E}$ is self-adjoint $({\cal E}^{\dag} = {\cal
E})$, then $f = {\cal S}^{\dag} (\psi)$ is a solution of ${\cal E} (f) =
0$. For example, in the case considered in the present work, the
(matrix) operator governing the field perturbations in the
Einstein-Maxwell-Dilaton theory is, in fact, self-adjoint \cite{12,13,17}.

Moreover, the existence of operators ${\cal S}$, ${\cal O}$, and ${\cal
T}$ satisfying the above identity, is equivalent to the existence of a
decoupled system
\[
   {\cal O} (\Psi) = 0, \nonumber
\]
obtained from the original system ${\cal E} (f) = 0$, such that the scalar
field $\Psi = {\cal T} (f)$.

Now, we can mix both branches of the adjoint operators scheme: since the
fields $\psi$ and $\Psi$ satisfy equation adjoints to each other, we can
establish, in according to the first branch, that
\[
   \psi ({\cal O} \Psi) - ({\cal O}^{\dag} \psi) \Psi = \nabla_{\mu}
{\cal J}^{\mu} (\psi, \Psi), \nonumber
\]
which means that $\nabla_{\mu} {\cal J}^{\mu} (\psi, \Psi) = 0$. 
Furthermore, since $\Psi$ is finally depending on $\psi$ $\left( \Psi = 
{\cal T} (f) = {\cal T} ({\cal S}^{\dag} (\psi)) \right)$, ${\cal
J}^{\mu}$ is dependent only on $\psi$ (however, see Section 5.4).

On the other hand, although this result on the existence of conserved
currents has been established assuming only tensor fields and the presence
of a single equation, such a result can be extended in a direct way to    
equations involving spinor fields, matrix fields, and the presence of more
than one field. Furthermore, this general result can be understood as an
important extension of the original Wald's method: wherever there exists
an appropriate decoupled equation, it is not only possible to express the
complete solution in terms of scalar potentials, but also to find
automatically a corresponding (covariantly) conserved current.  \\[2em]

\noindent {\uno III. BACKGROUND SPACETIME}
\vspace{1em}

Static, spherically symmetric solutions of the Einstein-Maxwell-dilaton
equations have been found, representing charged black holes for curvature 
below the Planck scale \cite{2,3}. The solutions for magnetically charged 
dilaton black holes have, using the metric convention (+ -- -- --), the 
line element
\begin{equation}
     ds^{2} = \chi^{2} dt^{2} - \chi^{-2} dr^{2} - R^{2} d \Omega,
\end{equation}
where $\chi$ and $R$ depend only on $r$:
\begin{eqnarray}
     \chi^{2} = \left( 1 - \frac{r_{+}}{r} \right) \left( 1 -
     \frac{r_{-}}{r} \right)^{(1 - a^{2})/ (1 + a^{2})}, \quad R = r
     \left( 1 - \frac{r_{-}}{r} \right)^{a^{2}/ (1 + a^{2})},
\end{eqnarray}
where $r_{+}$ and $r_{-}$ are the values of the parameter $r$ at the outer
and the inner horizon respectively, and are related to the physical mass
$(M)$ and charge $(Q)$; $a$ is the dilaton coupling parameter. The
Maxwell and dilaton fields are given by
\begin{eqnarray}
      F = Q {\rm sin} \theta d\theta \wedge d \varphi, \quad
      e^{-2 a \phi} = \left( 1 - \frac{r_{-}}{r} \right)^{2a^{2}/ (a^{2} + 
      1)}, \quad \left( \xi \equiv - e^{-2a \phi} \right) .
\end{eqnarray}
There are also electrically charged solutions which may be obtained by a
duality rotation. For more details see \cite{2,3}.

For our present purpose, it is more convenient to specify the line element
(2) by the null tetrad
\begin{eqnarray}
     & & D \equiv l^{\mu} \partial_{\mu} = \frac{1}{\chi^{2}} \partial_{t}
     + \partial_{r}, \qquad \qquad  \qquad  \Delta \equiv n^{\mu}  
     \partial_{\mu} = \frac{1}{2} (\partial_{t} - \chi^{2} \partial_{r}), 
     \nonumber \\
     & & \delta \equiv m^{\mu} \partial_{\mu} = \frac{1}{\sqrt{2}R}
     (\partial_{\theta} + i {\rm csc}\theta \partial_{\varphi}), \qquad
     \overline{\delta} \equiv \overline{m}^{\mu} \partial_{\mu} =
     \frac{1}{\sqrt{2}R} (\partial_{\theta} - i {\rm csc} \theta
     \partial_{\varphi}).
\end{eqnarray}
Using the commutation relations of the tetrad (5), the nonvanishing spin
coefficients can be conveniently expressed as
\begin{eqnarray}
     & & \rho = D \ {\rm ln} R^{-1}, \quad \mu = \Delta \ {\rm ln} R,
     \quad \gamma = \Delta \ {\rm ln} \chi^{-1}, \nonumber \\
     & & \beta = \delta \ {\rm ln} \ {\rm sin}^{1/2} \theta, \quad \alpha
     = - \overline{\delta} \ {\rm ln} \ {\rm sin}^{1/2} \theta,
\end{eqnarray}
where $\rho$, $\mu$, and $\gamma$ depend only on $r$, and $\beta$ and
$\alpha$ on both $r$ and $\theta$.

On the other hand, considering the first of Eqs.\ (4) and the definitions
$\varphi_{0} \equiv l^{\mu} m^{\nu} F_{\mu\nu}$, $\varphi_{1}  \equiv
\frac{1}{2} (l^{\mu} n^{\nu} + \overline{m}^{\mu} m^{\nu}) F_{\mu\nu}$,
and $\varphi_{2}  \equiv \overline{m}^{\mu} n^{\nu} F_{\mu\nu}$, the
Newman-Penrose components of the electromagnetic field are given by
\begin{equation}
     \varphi_{0} = 0 = \varphi_{2}, \quad \varphi_{1} (r) = 
     \frac{i Q}{2R^{2}}.
\end{equation}
Note that $\varphi_{1} + \overline{\varphi}_{1}= 0 = \delta \phi_{1}$, 
which will be used implicitly below. On the other hand, from Eqs.\ (4) and
(5), the only nonvanishing derivatives of the dilaton field are $D\phi$ 
and $\Delta\phi$, which depend only on $r$, and
\begin{equation}
     \delta \phi = 0 = \overline{\delta} \phi.
\end{equation}
Thus, the only nonvanishing Ricci scalars are (see Appendix of Ref.\
\cite{12})
\begin{eqnarray}
     \Phi_{00} = - (D\phi)^{2}, \quad \Phi_{22} = - (\Delta\phi)^{2},
     \nonumber \\
     \Phi_{11} = - \frac{1}{2} (D\phi) (\Delta\phi) - 2 \xi \varphi^{2}_{1},
     \quad \Lambda = - \frac{1}{6} (D\phi) (\Delta\phi),
\end{eqnarray}
and the only nonvanishing component of the Weyl spinor can be expressed as
\begin{equation}
     \Psi_{2}(r) = 2 \gamma \rho - \frac{2}{3} D\phi \Delta\phi.
\end{equation}
Furthermore, the background Maxwell's equations take the form \cite{12}
\begin{equation}
     (D - 2 \rho) \varphi_{1} = 0, \quad (\Delta + 2 \mu) \varphi_{1} = 0,
\end{equation}
and similarly, the background dilaton equation is
\begin{equation}
     D \Delta\phi + 2 \mu D\phi - 2 a \xi \varphi^{2}_{1} = 0.
\end{equation}
Additionally, using Eqs.\ (4)-(9) and the commutation relations, we can
find the following relations:
\begin{eqnarray}
     (D + p \rho) (\delta + q \beta) \!\! & = & \!\! (\delta + q \beta) [D
     + (p + 1) \rho], \nonumber \\
     (\Delta + p \gamma + p' \mu) (\delta + q \beta) \!\! & = & \!\! 
     (\delta + q \beta) [\Delta + p \gamma + (p' - 1) \mu],
\end{eqnarray}
where $p$, $q$, and $p'$ are three arbitrary constants.

In the Newman-Penrose formalism, the adjoints of the tetrad components
(5) are given, in general, by Eqs.\ (16) of Ref.\ \cite{12}, which reduce
to
\begin{eqnarray}
     D^{\dag} = - (D - 2 \rho), \quad \Delta^{\dag} = - (\Delta - 2 \gamma
     + 2 \mu), \quad \delta^{\dag} = - (\delta + 2 \beta), \quad
     \overline{\delta}^{\dag} = - (\overline{\delta} + 2\overline{\beta}),
\end{eqnarray}
for this background solution. These equations will be used below. \\[2em]

\noindent {\uno IV. DECOUPLED SET OF EQUATIONS FOR GAUGE INVARIANT
PERTURBATIONS}
\vspace{1em}

The notation, conventions, and Appendix of Ref.\ \cite{12} will be used
extensively throughout this paper. In particular, the metric, vector
potential, and dilaton variations are represented by $h_{\mu\nu}$,
$b_{\mu}$, and $\phi^{\rm B}$, respectively. The metric and vector
potential perturbations are defined modulo gauge transformations. Since, 
the dilaton is a fundamental physical field, there no exists gauge
invariance associated with this field.

On the other hand, it is well known that when the perturbation analysis is
performed using the Newman-Penrose formalism, one is faced with the
perturbed tetrad gauge freedom. The traditional approaches make use of
this gauge freedom in order to simplify the equations for the
perturbations (\cite{12} and references therein). However, we shall see
that in the present case, although including string fields, there is no
need to invoke perturbed tetrad rotations, but that appropriate 
combinations of the perturbed quantities, which are independent on the 
perturbed tetrad gauge freedom, lead in a natural way, to a decoupled set 
of equations from the original set. Such combinations prove to be also 
independent on the ordinary gauge transformations of the electromagnetic 
potential perturbations.

For example, let us consider the first-order perturbations of the spin
coefficient $\sigma$:
\begin{eqnarray}
     \sigma^{\rm B} \!\! & \equiv & \!\! - (l^{\mu} m^{\nu} \nabla_{\nu}
     m_{\mu})^{\rm B} = l^{\mu} m^{\nu} m_{\gamma}
     (\Gamma^{\gamma}_{\mu\nu})^{\rm B} - l^{\mu} m^{\nu} \nabla_{\nu}
     m^{\rm B}_{\mu} - (l^{\mu} m^{\nu})^{\rm B} \nabla_{\nu} m_{\mu}
     \nonumber \\
     \!\! & = & \!\! l^{\mu} m^{\nu} m_{\gamma} (\Gamma^{\gamma}_{\mu
     \nu})^{\rm B} - (\delta - 2 \beta) (l^{\mu} m^{\rm B}_{\mu}),
\end{eqnarray}
where it has been considered that the only nonvanishing spin coefficients
in the background are given in Eq.\ (6); $(\Gamma^{\gamma}_{\mu\nu})^{\rm 
B} = \frac{1}{2} g^{\gamma\rho} (\nabla_{\mu} h_{\nu\rho} + \nabla_{\nu}
h_{\mu\rho} - \nabla_{\rho} h_{\mu\nu})$, corresponds to the variations of
the connection, and in this manner, the first term in the above equation
is defined completely in terms of $h_{\mu\nu}$. On the order hand,
$l^{\mu} m^{\rm B}_{\mu}$ is dependent on the perturbed tetrad gauge 
freedom. Furthermore, from the definition $\varphi_{0} \equiv l^{\mu} 
m^{\nu} F_{\mu\nu}$, we have that
\begin{equation}
     \varphi^{\rm B}_{0} = l^{\mu} m^{\nu} F^{\rm B}_{\mu\nu} + 2
     \varphi_{1} (l^{\mu} m^{\rm B}_{\mu}),
\end{equation}
where Eq.\ (7) have been considered; $F^{\rm B}_{\mu\nu} = \partial_{\mu}
b_{\nu} - \partial_{\nu} b_{\mu}$, and thus the first term of Eq.\ (16) is
defined completely in terms of $b_{\mu}$. Therefore, from Eqs.\ (15), and
(16) we can see easily that the perturbed quantity $\tilde{\sigma}^{\rm B} 
\equiv \sigma^{\rm B} + (\delta - 2 \beta) (\varphi^{\rm B}_{0} / 2
\varphi_{1})$, is independent on the perturbed tetrad gauge freedom and 
defined completely in terms of $h_{\mu\nu}$ and $b_{\mu}$. Furthermore, 
since the field perturbation $F^{\rm B}_{\mu\nu}$ is invariant under the 
ordinary gauge transformation $b_{\mu} \rightarrow b_{\mu} + \nabla_{\mu} 
\varepsilon$, where $\varepsilon$ is an arbitrary scalar field, 
$\varphi^{\rm B}_{0}$ in Eq.\ (16) is also invariant under the 
transformation and, in this manner $\tilde{\sigma}^{\rm B} (h_{\mu\nu}, 
b_{\mu}) = \tilde{\sigma}^{\rm B} (h_{\mu\nu}, b_{\mu} + \nabla_{\mu} 
\varepsilon)$. The remaining quantities with similar invariance properties 
involved in our present analysis, are given in Appendix A.

For obtaining our first perturbation equation, we apply $(\delta - 2 \beta 
)$ to the first of Eqs.\ (A22), and using the commutation relations (13), 
we can use the first and second of Eqs.\ (A21), and first of Eq.\ (A23), 
for eliminating the resultant terms $(\delta - 2 \beta) \tilde{\kappa}^{\rm
B}$, $(\delta - 2 \beta) \tilde{\pi}^{\rm B}$, and $(\delta - 2 \beta)
\tilde{\Psi}^{\rm B}_{1}$ respectively, in favor of terms including
$\Psi^{\rm B}_{0}$, $\tilde{\sigma}^{\rm B}$, $\tilde{\lambda}^{\rm B}$,
and $\tilde{\phi}^{\rm B}$, and to obtain, after grouping suitably, the
second-order differential equation:
\begin{equation}
     {\cal O}_{11} \Psi^{\rm B}_{0} + {\cal O}_{13} \tilde{\sigma}^{\rm B}
     - (D\phi) F_{1} \tilde{\lambda}^{\rm B} + F_{1} (\delta - 2 \beta)
     \tilde{\phi}^{\rm B} = {\cal S}_{11} T_{\mu\nu},
\end{equation}
where
\begin{eqnarray}
     {\cal O}_{11} \!\! & = & \!\! (D - 5 \rho) (\Delta - 4 \gamma + \mu) 
     - (\delta - 2 \beta) (\overline{\delta} + 4 \overline{\beta}) - (3
     \Psi_{2} - 2 \Phi_{11} + 2 D\phi \Delta\phi), \nonumber \\
     {\cal O}_{13} \!\! & = & \!\! -8 \xi \varphi^{2}_{1} D - 4 D\phi
     (\gamma D\phi - 3 a \xi \varphi^{2}_{1}), \nonumber \\
     F_{1} (r) \!\! & = & \!\! 8 \chi^{-2} D\phi \left( \gamma +
     \frac{a \xi \varphi^{2}_{1}}{D\phi} \right),\\
\noalign{\hbox{and}}
     {\cal S}_{11} \!\! & = & \!\! 2 (\delta - 2 \beta) (D - 3 \rho) 
     l^{(\mu} m^{\nu)} - [(D - 5 \rho) (D - \rho) + \Phi_{00}] m^{\mu} 
     m^{\nu} - (\delta - 2 \beta) \delta l^{\mu} l^{\nu}. \nonumber \\
\end{eqnarray}

Similarly, applying $(\delta - 2 \beta)$ to the second of Eqs.\ (A22),
using the commutation relations (13), the fourth, fifth of Eqs.\ (A21),
and second of Eqs.\ (A23) for eliminating the resultant terms $(\delta - 2
\beta) \tilde{\tau}^{\rm B}$, $(\delta - 2 \beta) \tilde{\nu}^{\rm B}$,
and $(\delta - 2 \beta) \tilde{\Psi}^{\rm B}_{3}$, respectively, in favor
of terms involving $\overline{\Psi}^{\rm B}_{0}$, $\tilde{\sigma}^{\rm
B}$, $\tilde{\lambda}^{\rm B}$, and $\tilde{\phi}^{\rm B}$, one obtains 
another second-order differential equation:
\begin{equation}
     {\cal O}_{22} \overline{\Psi}^{\rm B}_{4} + \frac{\chi^{4}}{4} 
     \Delta\phi F_{1} \tilde{\sigma}^{\rm B} + {\cal O}_{24} 
     \tilde{\lambda}^{\rm B} +  \frac{\chi^{4}}{4}  F_{1} (\delta - 2 
     \beta) \tilde{\phi}^{\rm B} = {\cal S}_{21} T_{\mu\nu},
\end{equation}
where
\begin{eqnarray}
     {\cal O}_{22} \!\! & = & \!\! (\Delta + 2 \gamma + 5 \mu) (D - \rho) -
     (\delta - 2 \beta) (\overline{\delta} + 4 \overline{\beta}) - (3
     \Psi_{2} + 2 D\phi \Delta\phi - 2 \Phi_{11}), \nonumber \\
     {\cal O}_{24} \!\! & = & \!\! 8 \xi \varphi^{2}_{1} (\Delta + 2
     \gamma) + 4 \Delta\phi (\gamma D\phi - 3 a \xi \varphi^{2}_{1}), \\
\noalign{\hbox{and}}
     {\cal S}_{21} \!\! & = & \!\! 2 (\delta - 2 \beta) (\Delta + 2 \gamma
     + 3 \mu) n^{(\mu} m^{\nu)} - [(\Delta + 2 \gamma + 5 \mu) (\Delta +
     \mu) + \Phi_{22}] m^{\mu} m^{\nu} \nonumber \\
     \!\! & & \!\! - (\delta - 2 \beta) \delta n^{\mu} n^{\nu}.
\end{eqnarray}
With the purpose of obtaining perturbation equations which involve only
the perturbation quantities appearing in Eqs.\ (17) and (20), we 
substitute directly $\hat{\varphi}^{\rm B}_{1}$ and $\check{\varphi}^{\rm 
B}_{1}$ from Eqs.\ (A6) and (A7) respectively into Eq.\ (A12), and then 
substituting the resultant term $D \tilde{\tau}^{\rm B}$ from the third of 
Eqs. (A21), we
obtain:
\begin{eqnarray}
     \!\! & & \!\! - 2 \tilde{\Psi}^{\rm B}_{1} - (\Delta - 4 \gamma -
     \mu) \tilde{\kappa}^{\rm B} + (\overline{\delta} + 4
     \overline{\beta}) \tilde{\sigma}^{\rm B} + a D\phi (\tilde{\pi}^{\rm 
     B} + \tilde{\tau}^{\rm B}) - a (D - \rho) \tilde{\phi}^{\rm B} 
     \nonumber\\
     \!\! & & \!\! = \frac{1}{2 \varphi_{1}} [\delta (\xi^{-1} l^{\mu}
     j_{\mu}) - (D - 3 \rho) (\xi^{-1} m^{\mu} j_{\mu})],
\end{eqnarray}
further, applying $(\delta - 2 \beta)$ to the above equation, using the
commutation relations (13), and substituting the resultant terms $(\delta
- 2 \beta) \tilde{\Psi}^{\rm B}_{1}$, $(\delta - 2 \beta) \tilde{\kappa}^{\rm
B}$, $(\delta - 2 \beta) \tilde{\pi}^{\rm B}$, and $(\delta - 2 \beta)
\tilde{\tau}^{\rm B}$ from first of Eqs.\ (A23), first, second, and fourth
of Eqs.\ (A21) respectively, we obtain
\begin{equation}
     {\cal O}_{31} \Psi^{\rm B}_{0} + {\cal O}_{33} \tilde{\sigma}^{\rm B}
     + {\cal O}_{34} \tilde{\lambda}^{\rm B} + {\cal O}_{35} (\delta - 2 
     \beta) \tilde{\phi}^{\rm B} = {\cal S}_{31} (T_{\mu\nu}) + 
     {\cal S}_{32} (j_{\mu}),
\end{equation}
where
\begin{eqnarray}
     {\cal O}_{31} \!\! & = & \!\! \Delta - 4 \gamma + 2 \mu, \nonumber \\
     {\cal O}_{33} \!\! & = & \!\! (\Delta - 4 \gamma) (D - 2 \rho) - a D
\phi (\Delta - 2 \gamma) - (\delta - 2 \beta) (\overline{\delta} + 4
\overline{\beta}) - 2 (3 \Psi_{2} + 2 \Phi_{11}), \nonumber \\
     {\cal O}_{34} \!\! & = & \!\! - a D\phi D + 2 \Phi_{00}, \nonumber \\
     {\cal O}_{35} \!\! & = & \!\! a (D - 2 \rho) - 2 D\phi, \\
\noalign{\hbox{and}}
     {\cal S}_{31} \!\! & = & \!\! 2 (\delta - 2 \beta) l^{(\mu}m^{\nu)} -
2 (D - \rho) m^{\mu} m^{\nu}, \nonumber \\
     {\cal S}_{32} \!\! & = & \!\! \frac{1}{2 \varphi_{1}} (\delta - 2
\beta) [(D - 3 \rho) \xi^{-1} m^{\mu} - \delta \xi^{-1} l^{\mu}].
\end{eqnarray}

Similarly, following the above procedure for obtaining the equation (24),
we substitute  $\hat{\varphi}^{\rm B}_{1}$ and $\check{\varphi}^{\rm 
B}_{1}$ from Eqs.\ (A6) and (A7) into Eq.\ (A14), and then substituting 
the resultant term $\Delta \tilde{\pi}^{\rm B}$ from the sixth of Eqs.\ 
(A21), we obtain:
\begin{equation}
     2 \tilde{\Psi}^{\rm B}_{3} - (D + \rho) \tilde{\nu}^{\rm B} +
     (\overline{\delta} + 4 \overline{\beta}) \tilde{\lambda}^{\rm B} + a
     \Delta\phi (\tilde{\pi}^{\rm B} + \tilde{\tau}^{\rm B}) + a (\Delta 
     + \mu) \tilde{\phi}^{\rm B} = \frac{1}{2 \varphi_{1}} [(\Delta + 3 
     \mu) (\xi^{-1} m^{\mu} j_{\mu}) - \delta (\xi^{-1} n^{\mu} j_{\mu})],
\end{equation}
now, applying $(\delta - 2 \beta)$ to Eq.\ (27), using the commutation
relations (13), and substituting the resultant terms $(\delta - 2 \beta)
\tilde{\Psi}^{\rm B}_{3}$, $(\delta - 2 \beta) \tilde{\nu}^{\rm B}$,
$(\delta - 2 \beta) \tilde{\pi}^{\rm B}$, and $(\delta - 2 \beta)
\tilde{\tau}^{\rm B}$ from second of Eqs.\ (A23), fifth, second, and
fourth of Eqs.\ (A21), respectively, we obtain:
\begin{equation}
     {\cal O}_{42} \overline{\Psi}^{\rm B}_{4} + {\cal O}_{43}
     \tilde{\sigma}^{\rm B} + {\cal O}_{44} \tilde{\lambda}^{\rm B} +
     {\cal O}_{45} (\delta - 2 \beta) \tilde{\phi}^{\rm B} = {\cal S}_{41}
     (T_{\mu\nu}) + {\cal S}_{42} (j_{\mu}),
\end{equation}
where
\begin{eqnarray}
     {\cal O}_{42} \!\! & = & \!\! D - 2 \rho, \nonumber \\
     {\cal O}_{43} \!\! & = & \!\! a \Delta\phi (\Delta - 2 \gamma) - 2
\Phi_{22}, \nonumber \\
     {\cal O}_{44} \!\! & = & \!\! - D (\Delta + 2 \gamma + 2 \mu) + a
\Delta\phi D + (\delta - 2 \beta) (\overline{\delta} + 4 
\overline{\beta}) + 2 (3 \Psi_{2} + 2 \Phi_{11}), \nonumber \\
     {\cal O}_{45} \!\! & = & \!\! a (\Delta + 2 \mu) - 2 \Delta\phi, \\
\noalign{\hbox{and}}
     {\cal S}_{41} \!\! & = & \!\! 2 (\delta - 2 \beta) n^{(\mu} m^{\nu)}
- 2 (\Delta + \mu) m^{\mu} m^{\nu}, \nonumber \\
     {\cal S}_{42} \!\! & = & \!\! \frac{1}{2 \varphi_{1}} (\delta - 2
\beta) [(\Delta + 3 \mu) \xi^{-1} m^{\mu} - \delta \xi^{-1} n^{\mu}].
\end{eqnarray}

Similarly, substituting $\hat{\varphi_{1}}^{\rm B}$, and
$\check{\varphi_{1}}^{\rm B}$ from Eqs.\ (A6) and (A7) into Eq.\ (A20),
then applying $(\delta - 2 \beta)$ to the resultant equation (and
performing substitutions such as in the above equations for $(\delta - 2 
\beta) \tilde{\Psi}_{3}^{\rm B}$, $(\delta - 2 \beta) \tilde{\nu}^{\rm
B}$, $(\delta - 2 \beta) \tilde{\pi}^{\rm B}$, $(\delta - 2 \beta) 
\tilde{\tau}^{\rm B}$, and $(\delta - 2 \beta) \tilde{\kappa}^{\rm B})$, 
we obtain:
\begin{equation}
   \frac{\chi^{4}}{8}  F_{1}\Psi_{0}^{\rm B} + \frac{1}{2}F_{1} 
\overline{\Psi}_{4}^{\rm B} + {\cal O}_{53} \tilde{\sigma}^{\rm B} + {\cal 
O}_{54} \tilde{\lambda}^{\rm B} + {\cal O}_{55} (\delta - 2 \beta) 
\tilde{\phi}^{\rm B} = {\cal S}_{51}(T_{\mu\nu}) + {\cal S}_{52} (j_{\mu}) 
+ {\cal S}_{53} (\phi_{s}),
\end{equation}
where
\begin{eqnarray}
   {\cal O}_{53} \!\! & = & \!\! - \frac{\chi^{4}}{8} F_{1}(D - 2 \rho) + 
[\Delta\phi (D - \rho) - 2a \xi \varphi_{1}^{2}] (\Delta - 2\gamma +
\mu) - \Delta\phi (\delta - 2 \beta) (\overline{\delta} + 4
\overline{\beta}) \nonumber \\
   \!\! & & \!\! - \mu F_{2} - \Phi_{22}D\phi, \nonumber \\
   {\cal O}_{54} \!\! & = & \!\! (F_{2} - \mu D\phi)(D - \rho) - (D - 3
\rho) D\phi (\Delta + 2 \gamma + 2 \mu) + D\phi (\delta - 2 \beta)
(\overline{\delta} + 4 \overline{\beta}) \nonumber \\
   \!\! & & \!\! + [(\Delta\phi D - 2a \xi \varphi_{1}^{2})\rho] + D\phi
(3\Psi_{2} + 2\Phi_{11}), \nonumber\\
   {\cal O}_{55} \!\! & = & \!\! (D - 3 \rho)(\Delta + 3 \mu)- (\delta - 2
\beta) (\overline{\delta} + 4\overline{\beta}) - 3\Psi_{2} + 2 \mu \rho -
3 D\phi \Delta\phi - 4(a^{2} - 1) \xi \varphi_{1}^{2}, \nonumber \\
   F_{2} \!\! & \equiv & \!\! 2D\phi \left( \mu + \frac{a\xi
\varphi_{1}^{2}}{D\phi} \right) ,
\end{eqnarray}
and
\begin{eqnarray}
   {\cal S}_{51} \!\! & = & \!\! 2 D\phi (\delta - 2\beta) n^{(\nu}
m^{\nu)} + 2 \Delta\phi (\delta - 2 \beta) l^{(\nu} m^{\nu)} + (4a \xi 
\varphi_{1}^{2} - \Delta\phi D - D\phi \Delta) m^{\mu} m^{\nu},
\nonumber\\
   {\cal S}_{52} \!\! & = & \!\! - 4 a \varphi_{1} (\delta - 2 \beta) 
m^{\mu}, \nonumber\\
   {\cal S}_{53} \!\! & = & \!\! \frac{1}{2}(\delta - 2 \beta) \delta.
\end{eqnarray}
Hence, we have finally a system of five second-order linear partial
differential equations (17), (20), (24), (28), and (31), for five
unknowns: $\Psi_{0}^{\rm B}$, $\overline{\Psi}_{4}^{\rm B}$, 
$\tilde{\sigma}$, $\tilde{\lambda}$, and $(\delta - 2 \beta)
\tilde{\phi}^{\rm B}$ (in Ref.\ \cite{12}, a similar system was obtained 
for the equations governing the perturbations of the solution that 
represents waves bound to collisions in the same scheme of the 
Einstein-Maxwell-dilaton theory). This system of equations can be
expressed in the following matrix form:
\begin{equation}
    {\cal O} (\Psi^{\rm B})  = {\cal S} \pmatrix{ (T_{\mu\nu}) \cr
    (j_{\mu}) \cr \phi_{s} \cr},
\end{equation}
where ${\cal O}$ is the $5 \times 5$ matrix
\begin{equation}
    {\cal O} \equiv \pmatrix{{\cal O}_{11} & 0 & {\cal O}_{13} 
    & - F_{1} D\phi & F_{1} \cr 
    0 & {\cal O}_{22} & \frac{\chi^{4}}{4} \Delta\phi F_{1} 
    & {\cal O}_{24} & \frac{\chi^{4}}{4} F_{1} \cr
    {\cal O}_{31} & 0 & {\cal O}_{33} & {\cal O}_{34} & {\cal O}_{35} \cr
    0 & {\cal O}_{42} & {\cal O}_{43} & {\cal O}_{44} & {\cal O}_{45} \cr
    \frac{\chi^{4}}{8} F_{1} & \frac{1}{2} F_{1} & {\cal O}_{53} & 
    {\cal O}_{54} & {\cal O}_{55} \cr},
\end{equation}
\begin{equation}
    (\Psi^{\rm B}) \equiv \pmatrix{ \Psi_{0}^{\rm B} \cr
    \overline{\Psi}_{4}^{\rm B} \cr \tilde{\sigma}^{\rm B} \cr
    \tilde{\lambda}^{\rm B} \cr (\delta-2\beta) \tilde{\phi}^{\rm B} \cr},
\end{equation}
and ${\cal S}$ the $5 \times3$ matrix:
\begin{equation}
    {\cal S} \equiv \pmatrix{{\cal S}_{11} & 0 & 0 \cr
                    {\cal S}_{21} & 0 & 0 \cr
                    {\cal S}_{31} & {\cal S}_{32} & 0 \cr
                    {\cal S}_{41} & {\cal S}_{42} & 0 \cr
                    {\cal S}_{51} & {\cal S}_{52} & {\cal S}_{53} \cr}.
\end{equation}

Note that both ${\cal O}$ and ${\cal S}$ depend only on the background
fields. As mentioned previously, a gauge-fixing condition on the perturbed
tetrad is unnecessary for obtaining the complete system (34). Furthermore,
the entries of the matrix $(\Psi^{\rm B})$ are automatically independent
on the gauge transformations of the vector potential variations $b_{\mu}$ 
(see paragraph after Eq.\ (16)): $(\Psi^{\rm B}) (h_{\mu\nu}, b_{\mu}) =
(\Psi^{\rm B}) (h_{\mu\nu}, b_{\mu} + \nabla_{\mu} \varepsilon)$. In this 
manner, the invariance under the gauge freedoms of the matter fields and
the perturbed tetrad is guaranteed. This issue will be particularly 
important below, when we discuss the bilinear forms on the reduced phase 
space.

In the traditional approach, the field perturbations are separated in
polar and axial perturbations (and some gauge-fixing conditions are 
imposed) with the purpose of reducing the equations governing the 
perturbations to Schr\"{o}dinger-type equations, and then to apply 
semiclassical methods based on the Hermiticity of such system of
equations. However, as shown in Ref.\ \cite{18}, such treatment is 
unnecessary, and for many aims one can obtain essentially the same
physical results working directly with the original non-Hermitian system
of equations. In fact, when string fields are involved, such as the
present case, those reductions seem to be very difficult to carry out, or 
when possible, the interaction matrix is too complex to be displayed in 
explicit form \cite{5}. Therefore, Eqs.\ (34) in its original form, 
without separations nor reductions, are sufficient for our present 
purposes. \\[2em]

\noindent {\uno V. LOCAL CONTINUITY LAWS ON THE PHASE SPACE AND DEBYE
POTENTIALS}
\vspace{.5cm}

\noindent{\uno 5.1 Equations for the Debye potentials}
\vspace{.5cm}

Following the ideas of Section II (see for example that made in Ref.\
\cite{12}), if the matrix potential $(\psi)$ satisfies ${\cal O}^{\dag} 
(\psi) = 0$, with
\begin{equation}
   (\psi) = \pmatrix{ \psi_{G} \cr \psi_{H} \cr \psi_{E} \cr \psi_{F} \cr
   \psi_{D} \cr},
\end{equation}
then the metric, vector potential, and dilaton {\it real} variations are 
given by
\begin{eqnarray}
& &  \pmatrix{ -\frac{1}{2} h_{\mu\nu} \cr 2 b_{\mu} \cr \phi^{\rm B} \cr}
     = {\cal S}^{\dag} (\psi)
     = \pmatrix{{\cal S}_{11}^{\dag} & {\cal S}_{21}^{\dag} & {\cal
     S}_{31}^{\dag} & {\cal S}_{41}^{\dag} & {\cal S}_{51}^{\dag} \cr
     0 & 0 & {\cal S}_{32}^{\dag} & {\cal S}_{42}^{\dag} & {\cal
     S}_{52}^{\dag} \cr
     0 & 0 & 0 & 0 & {\cal S}_{53}^{\dag} \cr}
     \pmatrix{\psi_{\rm G} \cr \psi_{\rm H} \cr \psi_{\rm E} \cr
              \psi_{\rm F} \cr \psi_{\rm D} \cr}  \nonumber \\
& & = \pmatrix{ {\cal S}_{11}^{\dag}\psi_{G} + {\cal
S}_{21}^{\dag}\psi_{H} + {\cal S}_{31}^{\dag}\psi_{E} +
{\cal S}_{41}^{\dag}\psi_{F} + {\cal S}_{51}^{\dag}\psi_{D} + c.c \cr
{\cal S}_{32}^{\dag}\psi_{E} + {\cal S}_{42}^{\dag}\psi_{F} +
{\cal S}_{52}^{\dag}\psi_{D} + c.c \cr
{\cal S}_{53}^{\dag}\psi_{D} + c.c \cr};
\end{eqnarray}
from Eqs.\ (14), (19), (22), (26), (30), and (33) we have explicitly that,
\begin{eqnarray}
    {\cal S}_{11}^{\dag} \!\! & = & \!\! 2 l_{(\mu} m_{\nu)} (D + \rho) 
(\delta + 4 \beta) - m_{\mu} m_{\nu} [(D - \rho) (D + 3 \rho) + \Phi_{00}] 
- l_{\mu} l_{\nu} (\delta + 2 \beta) (\delta + 4 \beta), \nonumber \\
    {\cal S}_{21}^{\dag} \!\! & = & \!\! 2 n_{(\mu} m_{\nu)} (\Delta -
4 \gamma - \mu) (\delta + 4 \beta) - m_{\mu} m_{\nu} [(\Delta - 2 \gamma +
\mu) (\Delta - 4 \gamma - 3 \mu) + \Phi_{22}] \nonumber \\
\!\! & & \!\! - n_{\mu} n_{\nu} (\delta + 2 \beta) (\delta + 4 \beta),
\nonumber \\
    {\cal S}_{31}^{\dag} \!\! & = & \!\! 2 m_{\mu} m_{\nu} (D - \rho) - 2
l_{(\mu} m_{\nu)} (\delta + 4 \beta), \nonumber \\
    {\cal S}_{41}^{\dag} \!\! & = & \!\! 2 m_{\mu} m_{\nu} (\Delta - 2
\gamma + \mu) - 2 n_{(\mu} m_{\nu)} (\delta + 4 \beta), \nonumber \\
    {\cal S}_{51}^{\dag} \!\! & = & \!\! - 2 D\phi n_{(\mu} m_{\nu)}
(\delta + 4 \beta) - 2 \Delta\phi l_{(\mu} m_{\nu)} (\delta + 4\beta) + 
m_{\mu} m_{\nu} (8 a \xi \varphi_{1}^{2} + \Delta\phi D + D\phi \Delta), 
\nonumber \\
    {\cal S}_{32}^{\dag} \!\! & = & \!\! \frac{1}{2\xi} [m_{\mu} (D +
\rho) - l_{\mu} (\delta + 2 \beta)] (\delta + 4 \beta) 
\frac{1}{\varphi_{1}}, \nonumber \\
    {\cal S}_{42}^{\dag} \!\! & = & \!\! \frac{1}{2\xi} [m_{\mu} (\Delta 
- 2 \gamma - \mu) - n_{\mu} (\delta + 2 \beta)] (\delta + 4 \beta) 
\frac{1}{\varphi_{1}}, \nonumber \\
    {\cal S}_{52}^{\dag} \!\! & = & \!\! 4a \varphi_{1} m_{\mu} (\delta 
+ 4 \beta), \nonumber \\
    {\cal S}_{53}^{\dag} \!\! & = & \!\! \frac{1}{2} (\delta + 2 \beta) 
(\delta + 4 \beta).
\end{eqnarray}
In this manner, the complete field variations are given by Eqs.\ (39) in
terms of the Debye potentials, which satisfy a system of five second-order 
linear partial differential equations:
\begin{equation}
{\cal O}^{\dag}(\psi) = \pmatrix{{\cal O}_{11}^{\dag} & 0 & {\cal
            O}_{31}^{\dag} & 0 & \frac{\chi^{4}}{8} F_{1} \cr
            0 & {\cal O}_{22}^{\dag} & 0 & {\cal O}_{42}^{\dag} 
            & \frac{1}{2} F_{1} \cr
            {\cal O}_{13}^{\dag} & \frac{\chi^{4}}{4} \Delta\phi F_{1} 
            & {\cal O}_{33}^{\dag} & {\cal O}_{43}^{\dag} 
            & {\cal O}_{53}^{\dag} \cr
            - F_{1} D\phi & {\cal O}_{24}^{\dag} & {\cal O}_{34}^{\dag} 
            & {\cal O}_{44}^{\dag} & {\cal O}_{54}^{\dag} \cr
            F_{1} & \frac{\chi^{4}}{4} F_{1} & {\cal O}_{35}^{\dag} 
            & {\cal O}_{45}^{\dag} & {\cal O}_{55}^{\dag} \cr}
    \pmatrix{ \psi_{G} \cr \psi_{H} \cr \psi_{E} \cr
    \psi_{F} \cr \psi_{D} \cr} = 0,
\end{equation}
where
\begin{eqnarray}
{\cal O}_{11}^{\dag} \!\! & = & \!\! (\Delta + 2 \gamma + \mu) (D + 3 
     \rho) - (\overline{\delta} - 2 \overline{\beta}) (\delta + 4 \beta) 
     - (3 \Psi_{2} - 2 \Phi_{11} + 2 D\phi \Delta\phi), \nonumber \\
{\cal O}_{13}^{\dag} \!\! & = & \!\! 8 \xi \varphi^{2}_{1} (D + 2 \rho)
     + F_{1} \Delta\phi, \nonumber \\
{\cal O}_{22}^{\dag} \!\! & = & \!\! (D - \rho) (\Delta - 4 \gamma - 3
     \mu) - (\overline{\delta} - 2 \overline{\beta}) (\delta + 4 \beta) 
     - (3 \Psi_{2} - 2 \Phi_{11} + 2 D\phi \Delta\phi ), \nonumber \\
{\cal O}_{24}^{\dag} \!\! & = & \!\!  -8 \xi \varphi^{2}_{1} (\Delta - 4
     \gamma - 2 \mu) + \frac{\chi^{2}}{2} F_{1} \Delta\phi, \nonumber\\
{\cal O}_{31}^{\dag} \!\! & = & \!\! - (\Delta + 2 \gamma), \nonumber\\
{\cal O}_{33}^{\dag} \!\! & = & \!\! D (\Delta + 2 \gamma + 2 \mu) + a
     D\phi (\Delta + 2\mu) - (\overline{\delta} - 2\overline{\beta})
     (\delta + 4 \beta) - 2 (3 \Psi_{2} + 2 \Phi_{11}) + a \Delta D\phi,
     \nonumber\\
{\cal O}_{34}^{\dag} \!\! & = & \!\! a D\phi (D - 2 \rho) + a D^{2}\phi +
     2 \Phi_{00}, \nonumber \\
{\cal O}_{35}^{\dag} \!\! & = & \!\! - a D - 2 D\phi, \quad  {\cal
     O}_{42}^{\dag} = - D, \quad {\cal O}_{43}^{\dag} =  - a \Delta\phi
     (\Delta + 2 \mu) - a \Delta^{2}\phi - 2 \Phi_{22}, \nonumber \\
{\cal O}_{44}^{\dag} \!\! & = & \!\! - (\Delta - 4 \gamma + a \Delta\phi) 
     (D - 2 \rho) + (\overline{\delta} - 2 \overline{\beta}) (\delta +
     4 \beta) + 2 (3 \Psi_{2} + 2 \Phi_{11}) - a D \Delta\phi, \nonumber\\
{\cal O}_{45}^{\dag} \!\! & = & \!\! - a (\Delta - 2 \gamma) - 2
     \Delta\phi, \nonumber \\
{\cal O}_{53}^{\dag} \!\! & = & \!\! \frac{1}{8} D \chi^{4} F_{1} + 
     (\Delta + \mu) (4a \xi \varphi^{2}_{1} - \mu D\phi + \Delta\phi D) 
     - \Delta\phi (\overline{\delta} - 2 \overline{\beta}) (\delta + 
     4 \beta) - \Phi_{22} D\phi - \mu F_{2}, \nonumber \\
{\cal O}_{54}^{\dag} \!\! & = & \!\! - (D - \rho) (F_{2} - \mu D\phi) - 
      (\Delta - 4 \gamma) D\phi (D + \rho) + D\phi (\overline{\delta} - 2 
      \overline{\beta}) (\delta + 4 \beta) \nonumber \\
\!\! & & \!\! + [(\Delta\phi D - 2a \xi \varphi_{1}^{2}) \rho] + D\phi
      (3 \Psi_{2} + 2 \Phi_{11}), \\
{\cal O}_{55}^{\dag} \!\! & = & \!\! (\Delta - 2 \gamma - \mu) (D + \rho)
      - (\overline{\delta} - 2 \overline{\beta}) (\delta + 4 \beta) -
      3 \Psi_{2} + 2 \mu \rho - 3 D\phi \Delta\phi + 4 (a^{2} - 1) \xi
      \varphi_{1}^{2}, \nonumber
\end{eqnarray}
and Eqs.\ (14), (18), (21), (25), (29), and (32) have been used. Eqs.\
(41) are our fundamental equations since, as we shall see, all conserved 
quantities and bilinear forms on the phase space are defined in terms of 
the Debye potentials. Although these equations admit separable solutions 
in a simple way, we will use them first in the form (41) in order to 
establish a {\it covariant} conservation law, and subsequently to carry
out such separation. \\[2em]

\noindent {\uno 5.2 Covariant continuity equation and bilinear forms on
the phase space}
\vspace{.5cm}

Since the decoupled system and the system of equations for the Debye 
potentials are adjoints to each other, in according to the results of 
Section II we have that
\begin{equation}
    (\psi) \wedge {\cal O}(\Psi^{\rm B}) - {\cal O}^{\dag}(\psi) \wedge
    (\Psi^{\rm B}) = \nabla_{\mu} J^{\mu} (\psi, \Psi^{\rm B}).
\end{equation}
The left-hand side contains terms of the form $\psi_{G} \wedge {\cal
O}_{11} \Psi^{\rm B}_{0} - {\cal O}^{\dag}_{11} \psi_{G} \wedge 
\Psi^{\rm B}_{0}$ (see Eqs.\ (35), and (41)), which can be expressed in
the following form, considering the explicit forms of the operators ${\cal 
O}_{11}$, and ${\cal O}_{11}^{\dag}$ given in Eqs.\ (18), and (42) 
respectively, that $D \equiv l^{\mu} \partial_{\mu}$, $\Delta \equiv
n^{\mu} \partial_{\mu}$, $\delta \equiv m^{\mu} \partial_{\mu}$, 
$\overline{\delta} \equiv \overline{m}^{\mu} \partial_{\mu}$, and that
they are acting on scalar fields:
\begin{eqnarray}
& & \psi_{G} \wedge {\cal O}_{11} \Psi^{\rm B}_{0} - {\cal O}_{11}^{\dag}
\psi_{G} \wedge \Psi^{\rm B}_{0} = \nabla_{\mu} [l^{\mu} \psi_{G} \wedge
(\Delta - 4\gamma + \mu) \Psi_{0}^{\rm B} - n^{\mu} (D + 3 \rho)
\psi_{G} \wedge \Psi_{0}^{\rm B}  \nonumber \\
& & - m^{\mu} \psi_{G} \wedge (\overline{\delta} + 4 \overline{\beta})
\Psi_{0}^{\rm B} + \overline{m}^{\mu} (\delta + 4 \beta) \psi_{G} \wedge
\Psi_{0}^{\rm B}],
\end{eqnarray}
and similarly for the remaining terms:
\begin{eqnarray}
\!\! & & \!\! \psi_{G} \wedge {\cal O}_{13} \tilde{\sigma}^{\rm B} - {\cal
      O}_{13}^{\dag} \psi_{G} \wedge \tilde{\sigma}^{\rm B} = \nabla_{\mu}
      (- 8\xi \varphi_{1}^{2}l^{\mu}\psi_{G} \wedge \tilde{\sigma}^{\rm
      B}), \nonumber\\
\!\! & & \!\! \psi_{H} \wedge {\cal O}_{22} \overline{\Psi}^{\rm B}_{4} -
      {\cal O}^{\dag}_{22} \psi_{H} \wedge \overline{\Psi}^{\rm B}_{4} =
      \nabla_{\mu} [n^{\mu} \psi_{H} \wedge (D - \rho) 
      \overline{\Psi}^{\rm B}_{4} - l^{\mu} (\Delta - 4 \gamma - 3 \mu) 
      \psi_{H} \wedge \overline{\Psi}^{\rm B}_{4} \nonumber \\
\!\! & & \!\! - m^{\mu} \psi_{H} \wedge (\overline{\delta} + 4
      \overline{\beta}) \overline{\Psi}^{\rm B}_{4} + \overline{m}^{\mu}
      (\delta + 4 \beta) \psi_{H} \wedge \overline{\Psi}^{\rm B}_{4}],
      \nonumber \\
\!\! & & \!\! \psi_{H} \wedge {\cal O}_{24} \tilde{\lambda}^{\rm B} -
      {\cal O}^{\dag}_{24} \psi_{H} \wedge \tilde{\lambda}^{\rm B} =
      \nabla_{\mu} [8 \xi \varphi_{1}^{2} n^{\mu} \psi_{H} \wedge
      \tilde{\lambda}^{\rm B}), \nonumber\\
\!\! & & \!\! \psi_{E} \wedge {\cal O}_{31} \Psi^{\rm B}_{0} - {\cal
      O}^{\dag}_{31} \psi_{E} \wedge \Psi^{\rm B}_{0} = \nabla_{\mu}
      [n^{\mu} \psi_{E} \wedge \Psi^{\rm B}_{0}], \nonumber \\
\!\! & & \!\! \psi_{E} \wedge {\cal O}_{33} \tilde{\sigma}^{\rm B} - {\cal
      O}^{\dag}_{33} \psi_{E} \wedge \tilde{\sigma}^{\rm B} = \nabla_{\mu}
      [n^{\mu} \psi_{E} \wedge (D - 2 \rho - a D\phi) \tilde{\sigma}^{\rm
      B} - l^{\mu} (\Delta + 2 \gamma + 2 \mu) \psi_{E} \wedge
      \tilde{\sigma}^{\rm B} \nonumber \\
\!\! & & \!\! - m^{\mu} \psi_{E} \wedge (\overline{\delta} + 4
      \overline{\beta}) \tilde{\sigma}^{\rm B} + \overline{m}^{\mu} 
      (\delta + 4 \beta) \psi_{E} \wedge \tilde{\sigma}^{\rm B}], 
      \nonumber \\
\!\! & & \!\! \psi_{E}\wedge {\cal O}_{34}\tilde{\lambda}^{\rm B} - {\cal
      O}^{\dag}_{34} \psi_{E} \wedge \tilde{\lambda}^{\rm B} =
      \nabla_{\mu} (-a D\phi l^{\mu} \psi_{E} \wedge 
      \tilde{\lambda}^{\rm B}), \nonumber\\
\!\! & & \!\! \psi_{E} \wedge {\cal O}_{35} (\delta - 2 \beta)
      \tilde{\phi}^{\rm B} - {\cal O}^{\dag}_{35} \psi_{E} \wedge (\delta 
      - 2 \beta) \tilde{\phi}^{\rm B} = \nabla_{\mu} [al^{\mu} \psi_{E}
      \wedge (\delta - 2 \beta) \tilde{\phi}^{\rm B}], \nonumber\\
\!\! & & \!\! \psi_{F} \wedge {\cal O}_{42} \overline{\Psi}^{\rm B}_{4} -
      {\cal O}^{\dag}_{42} \psi_{F} \wedge \overline{\Psi}^{\rm B}_{4} =
      \nabla_{\mu}(l^{\mu} \psi_{F} \wedge \overline{\Psi}^{\rm B}_{4}),
      \nonumber \\
\!\! & & \!\! \psi_{F} \wedge {\cal O}_{43} \tilde{\sigma}^{\rm B} - 
      {\cal O}^{\dag}_{43} \psi_{E} \wedge \tilde{\sigma}^{\rm B}  =
      \nabla_{\mu} [a \Delta\phi n^{\mu} \psi_{F} \wedge 
      \tilde{\sigma}^{\rm B}], \nonumber\\
\!\! & & \!\! \psi_{F}\wedge {\cal O}_{44} \tilde{\lambda}^{\rm B} - {\cal
      O}^{\dag}_{44} \psi_{F} \wedge \tilde{\lambda}^{\rm B} =
      \nabla_{\mu} [n^{\mu} (D - 2 \rho) \psi_{F} \wedge 
      \tilde{\lambda}^{\rm B} - l^{\mu} \psi_{F} \wedge (\Delta + 2 \gamma 
      + 2 \mu - a \Delta\phi) \tilde{\lambda}^{\rm B} \nonumber \\
\!\! & & \!\! + m^{\mu} \psi_{F} \wedge (\overline{\delta}
      + 4 \overline{\beta}) \tilde{\lambda}^{\rm B} - \overline{m}^{\mu}
      (\delta + 4 \beta) \psi_{F} \wedge \tilde{\lambda}^{\rm B}],
      \nonumber \\
\!\! & & \!\! \psi_{F} \wedge {\cal O}_{45} (\delta - 2 \beta) 
      \tilde{\phi}^{\rm B} - {\cal O}^{\dag}_{45} \psi_{F} \wedge (\delta 
      - 2 \beta) \tilde{\phi}^{\rm B} = \nabla_{\mu} [a n^{\mu} \psi_{F} 
      \wedge (\delta - 2 \beta) \tilde{\phi}^{\rm B}],\nonumber\\
\!\! & & \!\! \psi_{D} \wedge {\cal O}_{53}\tilde{\sigma}^{\rm B} - {\cal
      O}^{\dag}_{53} \psi_{D} \wedge \tilde{\sigma}^{\rm B} =
      \nabla_{\mu} [-n^{\mu} (\Delta\phi D - \mu D\phi + 4a \xi 
      \varphi_{1}^{2}) \psi_{D} \wedge \tilde{\sigma}^{\rm B} \nonumber \\
\!\! & & \!\! + l^{\mu} \Delta\phi \psi_{D} \wedge \left( \Delta + \mu +
      \frac{2a\xi \varphi_{1}^{2}}{D\phi} \right) \tilde{\sigma}^{\rm B}
      - \Delta\phi m^{\mu} \psi_{D} \wedge (\overline{\delta} + 4
      \overline{\beta}) \tilde{\sigma}^{\rm B} + \Delta\phi
      \overline{m}^{\mu}(\delta + 4 \beta) \psi_{D} \wedge 
      \tilde{\sigma}^{\rm B}], \nonumber \\
\!\! & & \!\! \psi_{D} \wedge {\cal O}_{54} \tilde{\lambda}^{\rm B} -
      {\cal O}^{\dag}_{54} \psi_{D} \wedge \tilde{\lambda}^{\rm B} =
      \nabla_{\mu} [D\phi n^{\mu} (D + \rho) \psi_{D} \wedge 
      \tilde{\lambda}^{\rm B} - D\phi l^{\mu} \psi_{D} \wedge \left( 
      \Delta + 2 \gamma + \mu - \frac{2a\xi \varphi_{1}^{2}}{D\phi} 
      \right) \tilde{\lambda}^{\rm B} \nonumber \\
\!\! & & \!\! + D\phi m^{\mu} \psi_{D} \wedge (\overline{\delta} + 4
      \overline{\beta}) \tilde{\lambda}^{\rm B} - D\phi \overline{m}^{\mu}
      (\delta + 4 \beta) \psi_{D} \wedge \tilde{\lambda}^{\rm B}],
      \nonumber \\
\!\! & & \!\! \psi_{D} \wedge {\cal O}_{55}(\delta - 2 \beta)
      \tilde{\phi}^{\rm B} - {\cal O}^{\dag}_{55} \psi_{D} \wedge (\delta
      - 2 \beta) \tilde{\phi}^{\rm B} = \nabla_{\mu} [-n^{\mu} (D + \rho)
      \psi_{D} \wedge (\delta - 2 \beta)\tilde{\phi}^{\rm B} \nonumber \\
\!\! & & \!\! + l^{\mu} \psi_{D} \wedge (\Delta + 3 \mu) (\delta - 2
      \beta) \tilde{\phi}^{\rm B} - m^{\mu} \psi_{D} \wedge 
      (\overline{\delta} + 4 \overline{\beta}) (\delta - 2 \beta) 
      \tilde{\phi}^{\rm B} \nonumber \\
\!\! & & \!\! + \overline{m}^{\mu} (\delta + 4 \beta) \psi_{D} \wedge
      (\delta - 2 \beta) \tilde{\phi}^{\rm B}].
\end{eqnarray}

Moreover, from Eqs.\ (34), and (41) ${\cal O}(\Psi^{\rm B}) = 0$
\footnote{The presence of an inhomogeneous term corresponding to the 
additional sources of the field variations in Eqs.\ (34), is only a {\it
knack} for finding the operator ${\cal S}$. Finally we set $T_{\mu\nu} = 
0$, $j_{\mu} = 0$, $\phi_{s} = 0$.}, and ${\cal O}^{\dag}(\psi) = 0$; 
hence, from Eq.\ (43) we have the local continuity law:
\begin{eqnarray}
\!\! & & \!\! \nabla_{\mu} J^{\mu} (\Psi^{\rm B}, \psi) = 0, \\
\!\! & & \!\! J^{\mu}= J^{\mu}_{11}+J^{\mu}_{13}+J^{\mu}_{22}+J^{\mu}_{24}
     +J^{\mu}_{31}+J^{\mu}_{33}+J^{\mu}_{34}+J^{\mu}_{35}+J^{\mu}_{42}
     +J^{\mu}_{43}+J^{\mu}_{44}+J^{\mu}_{45}+J^{\mu}_{53}+J^{\mu}_{54}
     +J^{\mu}_{55}, \nonumber
\end{eqnarray}
and, of course, the $J^{\mu}_{ij}$ 's $(i,j = 1,2,3,4,5)$ are the
components coming from Eqs.\ (44), and (45); for example, $J^{\mu}_{34} = 
- a D\phi l^{\mu} \psi_{E} \wedge \tilde{\lambda}^{\rm B}$. Thus,
$J^{\mu}$ is a {\it covariantly} conserved current. We will discuss now
the properties and physical meaning of $J^{\mu}$.

It is easy to verify that, such as $(\Psi^{\rm B})$ in Eq.\ (36), the
matrix potential $(\psi)$ in Eq.\ (38) is made out of one-forms. Eqs.\
(39) give the field variations $h_{\mu\nu}$, $b_{\mu}$, and $\phi^{\rm B}$ 
(one-forms), in terms of $(\psi)$. Since the operator ${\cal S}^{\dag}$ is 
dependent only on background fields (zero-forms), thus $(\psi)$
corresponds to one-forms. This implies automatically that $J^{\mu} =
J^{\mu} (\Psi^{\rm B}, \psi)$ in Eq.\ (46) is a (non-degenerate) two-form 
on the corresponding phase space of the solution considered (the matrix 
operators ${\cal O}$ and ${\cal O}^{\dag}$ involved in the construction
of $J^{\mu}$ are also dependent only on the background fields). In next 
section, we will demonstrate that $J^{\mu}$ is a {\it closed} two-form on
the phase space, from which a symplectic structure will be constructed. 
\\[2em]

\noindent {\uno 5.3 Covariant symplectic structure on the phase space}
\vspace{.5cm}

For demonstrating that $J^{\mu}$ is a closed two-form, we need rewrite
the $J^{\mu}_{ij}$ 's in Eq.\ (46). For example, $J^{\mu}_{11}$ (see Eq.\
(44)) can be rewritten as:
\begin{eqnarray}
\!\!\! & & \!\!\! l^{\mu} \psi_{G} \wedge (\Delta - 4 \gamma + \mu) 
      \Psi^{\rm B}_{0} - n^{\mu} (D + 3 \rho) \psi_{G} \wedge \Psi^{\rm 
      B}_{0} - m^{\mu} \psi_{G} \wedge (\overline{\delta} + 4 
      \overline{\beta}) \Psi^{\rm B}_{0} + \overline{m}^{\mu} (\delta + 
      4 \beta) \psi_{G} \wedge \Psi^{\rm B}_{0} \nonumber \\
\!\!\! & & \!\!\! = - [l^{\mu} \psi_{G} (\Delta - 4 \gamma + \mu)
      \Psi_{0}]^{\rm B} + [n^{\mu} (D + 3 \rho) \psi_{G} \Psi_{0}]^{\rm B}
      + [m^{\mu} \psi_{G} (\overline{\delta} + 4 \overline{\beta})
      \Psi_{0}]^{\rm B} - [\overline{m}^{\mu} (\delta + 4 \beta) \psi_{G}
      \Psi_{0}]^{\rm B}, \nonumber \\
\end{eqnarray}
where we have considered that $\Psi_{0}$ vanishes at the background, and
the Leibniz rule for the exterior derivative. Eq.\ (47) implies that  
$J^{\mu}_{11}$ is an {\it exact} two-form, and automatically a closed
two-form. Similarly, using the fact that $\overline{\Psi}^{\rm B}_{4},
\tilde{\sigma}^{\rm B}, \tilde{\lambda}^{\rm B}$, and $(\delta - 2
\beta) \tilde{\phi}^{\rm B}$ can be expressed as variations of vanishing  
background fields, and the property of exterior derivative used above, we
can find that:
\begin{equation}
      (J^{\mu}_{ij})^{\rm B} = 0, 
\end{equation}
which makes that $J^{\mu}$ itself to be closed. In this manner, the
geometrical structure defined as $\omega \equiv \int_{\Sigma} J^{\mu} d   
\Sigma_{\mu}$, where $\Sigma$ is an initial value hypersurface,
corresponds to a symplectic structure on the phase space. As $J^{\mu}$ is
conserved, $\omega$ is independent of the choice of $\Sigma$ and, in
particular, is Poincar\'e invariant. Since $(\Psi^{\rm B})$ is invariant 
under gauge transformation of $b_{\mu}$ (see paragraph after Eq.\ (16)),
$J^{\mu}$ and $\omega$ have the same invariance properties. Hence, we have
constructed a gauge-invariant closed two-form $\omega$ on the reduced
phase space, which means the phase space modulo gauge transformations.
Similarly, $J^{\mu}$ and $\omega$ are independent of the perturbed tetrad
gauge freedom. \\[2em]

\noindent {\uno 5.4 Debye potentials as fundamental geometrical
structures}
\vspace{.5cm}

As we have seen, the bilinear forms $J^{\mu}$ and $\omega$ depend on the
background fields and the solutions admitted by the decoupled system for
$(\Psi^{\rm B})$ and its adjoint system for the Debye potentials. However, 
the components of $(\Psi^{\rm B})$, as described in the Appendix A, are 
defined completely in terms of the field variations $h_{\mu\nu}$,
$b_{\mu}$, and $\phi^{\rm B}$, which in turn, are defined in terms of the 
Debye potentials (see Eqs.\ (39)). Therefore, $J^{\mu}$ and $\omega$ can 
be expressed finally in terms of a single solution of the equations for 
Debye potentials. However, in the more general case, if $(\psi)_{1}$ is a 
solution admitted by the equations for the potentials, the matrix 
$(\Psi^{\rm B})$ can be expressed in terms of a second solution 
$(\psi)_{2}$, in general different of $(\psi)_{1}$, and thus, $J^{\mu}$ 
and $\omega$ are defined in terms of a pair of solutions for those
equations. Therefore, the Debye potentials, which correspond to one-forms
on the phase space, become the fundamental geometrical objects. The 
analysis of the structure of the phase space (and the perturbation 
analysis) has been reduced to the study of scalar equations for the 
potentials, which is a relatively simple issue. As we will see below, 
conserved quantities will be also expressed completely in terms of the
same potentials. \\[2em]
% Additionally, it will be discussed the possible connection 
%between the Debye potentials and other fundamental objects of the 
%symplectic geometry: {\it the symplectic potentials}. \\[2em]

\noindent {\uno VI. SEPARATION OF VARIABLES AND CONSERVED QUANTITIES}
\vspace{.5cm}

Our fundamental equations for the Debye potentials (41) and the continuity
equation (46), admit separation of variables in terms of harmonic time and
the spin-weighted spherical harmonics. The first ones are reduced to a 
system of {\it ordinary} differential equations for the radial parts of
the potentials, the second one yields two conserved quantities expressed 
in terms of such radial parts. \\[2em]

\noindent {\uno 6.1 Separable solutions for the potentials}

An advantage of using the Newman-Penrose formalism is that each quantity
has a type, and its corresponding boost weight and spin weight. This
property suggests the separable solutions more convenient for the
equations under study.

More specifically, if $\eta$ is a quantity of type $\{ p,q\}$, the effect
of the (relevant) Geroch-Held-Penrose operators on $\eta$ is given by
$\partial \hspace{-0.2cm}/ \eta \equiv (\delta - p \beta - q 
\overline{\alpha}) \eta$, and $\partial\hspace{-0.2cm}/ ' \eta \equiv 
(\overline{\delta} - p \alpha - q \overline{\beta}) \eta$, which, using 
Eqs.\ (5) and (6), reduce to \cite{20}
\begin{eqnarray}
\partial \hspace{-0.2cm}/ \eta \!\! & = & \!\! \frac{\sin^{s}
    \theta}{\sqrt{2}R} (\partial_{\theta} + i \csc \theta 
    \partial_{\varphi}) \sin^{-s} \theta \eta, \nonumber \\
\partial\hspace{-0.2cm}/ ' \eta \!\! & = & \!\! \frac{\sin^{-s}
    \theta}{\sqrt{2}R} (\partial_{\theta} - i \csc \theta 
    \partial_{\varphi}) \sin^{s} \theta \eta,
\end{eqnarray}
where $s \equiv(p-q)/2$ is the spin weight of $\eta$. In the particular
case that $\eta = {_{s}}Y_{lm}$, which means the spin-weighted spherical 
harmonics:
\begin{eqnarray}
\partial\hspace{-0.2cm}/ {_{s}}Y_{lm} \!\! & = & \!\! (\delta - 2
    s\beta) \, {_{s}}Y_{lm} = \frac{1}{\sqrt{2}R} [(l-s)(l+s+1)]^{1/2} \,
    {_{s+1}}Y_{lm}, \nonumber \\
\partial\hspace{-0.19cm}/ ' {_{s}}Y_{lm} \!\! & = & \!\! 
    (\overline{\delta} + 2 s \overline{\beta}) \,  {_{s}}Y_{lm} = -
    \frac{1}{\sqrt{2}R} [(l+s)(l-s+1)]^{1/2} \, {_{s-1}}Y_{lm}.
\end{eqnarray}
On the other hand, from Eqs.\ (41), it is easy to determine that the
potentials $\psi_{G}$, $\psi_{H} $, $\psi_{E}$, $\psi_{F}$, and $\psi_{D}$ 
have types $\{ -4,0\}$, $\{ 0,4\}$, $\{ -3,1\}$, $\{ -1,3\}$, and 
$\{ -2,2\}$ respectively. Therefore, all potentials have spin weight --2.

Making use of the fact that the background solution is static and
spherically symmetric, we seek for solutions for the potentials of the 
form:
\begin{equation}
    \psi_{I} = \psi_{i}(r) \, {_{-2}}Y_{lm} \, (\theta, \varphi)
    e^{-i\omega t},
\end{equation}
where the subscript $I= G, H, E, F, D$, and $i=g, h, e, f, d$
respectively. Since $(\overline{\delta} - 2 \overline{\beta}) (\delta 
+ 4 \beta)$ is the only operator appearing in Eqs.\ (41), and (42) that 
involves angular variables, we only need to know that:
\begin{equation}
(\overline{\delta} - 2 \overline{\beta}) (\delta + 4 \beta) \psi_{I} =
      - \frac{L^{2}}{2R^{2}}\psi_{I}, \quad L = [(l-1)(l+2)]^{1/2},
\end{equation}
where Eqs.\ (50), and (51) have been employed. The remaining terms
correspond to functions and differential operators involving only radial
and time variables. In fact, from Eqs.\ (5), and (51) we have that:
\begin{eqnarray}
     & & D \psi_{I} = {\cal D} \psi_{I}, \quad \Delta \psi_{I} = -
\frac{\chi^{2}}{2} \overline{{\cal D}} \psi_{I}, \quad (D
\overline{\psi}_{I} = \overline{{\cal D}} \, \overline{\psi}_{I}, \quad
\Delta \overline{\psi}_{I} = - \frac{\chi^{2}}{2} {\cal D}
\overline{\psi}_{I}), \\
\noalign{\hbox{where}}
     & & {\cal D} = \partial_{r} - \frac{i\omega}{\chi^{2}}, \quad
\overline{{\cal D}} = \partial_{r} + \frac{i\omega}{\chi^{2}}.
\end{eqnarray}

In this manner, it suffices to substitute the operators $D$ and $\Delta$,
in according to Eqs.\ (53), by ${\cal D}$ and $- \frac{\chi^{2}}{2} 
\overline{{\cal D}}$ respectively, $(\overline{\delta} - 2
\overline{\beta}) (\delta + 4 \beta)$ by $- \frac{L^{2}}{2 R^{2}}$ (in 
according to Eq.\ (52)), and $\psi_{I}$ by $\psi_{i}$ (the corresponding 
radial part) into Eqs.\ (41), for reducing them to an system of ordinary 
equations for the radial parts $\psi_{i}$'s of the potentials. Hence, the 
separation of variables proposed in Eq.\ (51) applies in a natural and 
straightforward way. \\[2em]

\noindent {\uno 6.2 Separation of variables for the continuity equation}
\vspace{2em}

In this section we will see that the covariant continuity equation (46), 
together with the separable solutions admitted for the potentials (Eq.\
(51)), and the corresponding separation of variables for the field
variations (Appendix B), lead to the existence of two conserved
quantities.

As we have seen, at each spacetime point, $J^{\mu}$ in Eq.\ (46) is a
two-form on the phase space. Regardless of the last interpretation, we can
maintain $J^{\mu}$ as a bilinear product on field perturbations on the
spacetime manifold. In this manner, the covariantly conserved current
(46) can be rewritten, grouping conveniently its components on the null 
tetrad, in the form:
\begin{equation}
   J^{\mu} = V_{l} l^{\mu} + V_{n} n^{\mu} + V_{m} m^{\mu} +
   V_{\overline{m}} \overline{m}^{\mu},
\end{equation}
where
\begin{eqnarray}
V_{l} \!\! & \equiv & \!\! \psi_{\rm G} (\Delta - 4 \gamma + \mu) 
     \Psi^{\rm B}_{0} - 8 \xi \varphi^{2}_{1} \psi_{\rm G} 
     \tilde{\sigma}^{\rm B} - \overline{\Psi}^{\rm B}_{4} (\Delta - 4 
     \gamma - 3 \mu) \psi_{H} - \tilde{\sigma}^{\rm B} (\Delta + 2 \gamma 
     + 2 \mu) \psi_{\rm E} \nonumber \\
     \!\! & & \!\! - a D\phi  \psi_{E} \tilde{\lambda}^{\rm B} + a
     \psi_{E} (\delta - 2 \beta) \tilde{\phi}^{\rm B} + \psi_{\rm F}
     \overline{\Psi}^{\rm B}_{4} - \psi_{\rm F} (\Delta + 2 \gamma + 2 \mu
     - a \Delta\phi) \tilde{\lambda}^{\rm B} \nonumber \\
     \!\! & & \!\! + \Delta\phi \psi_{d} \left( \Delta + \mu +
     \frac{2a\xi\varphi^{2}_{1}}{D\phi} \right) \tilde{\sigma}^{\rm B} -
     D\phi \psi_{d} \left( \Delta + 2 \gamma + \mu - \frac{2a\xi
     \varphi^{2}_{1}}{D\phi} \right) \tilde{\lambda}^{\rm B} \nonumber \\
     & & + \psi_{\rm D} (\Delta + 3 \mu)] (\delta - 2 \beta)
     \tilde{\phi}^{\rm B}, \nonumber \\
V_{n} \!\! & \equiv & \!\! - \Psi^{\rm B}_{0} (D + 3 \rho) \psi_{\rm G} +
     \psi_{\rm H} (D - \rho) \overline{\Psi}^{\rm B}_{4} + 8 \xi
     \varphi^{2}_{1} \psi_{\rm H} \tilde{\lambda}^{\rm B} + \psi_{\rm E}
     \Psi^{\rm B}_{0} + \psi_{\rm E} (D - 2 \rho - a D\phi)
     \tilde{\sigma}^{\rm B} \nonumber \\
     \!\! & & \!\! + a \Delta\phi \psi_{\rm F} \tilde{\sigma}^{\rm B}
     + \tilde{\lambda}^{\rm B} (D - 2 \rho) \psi_{\rm F}
     + a \psi_{\rm F} (\delta - 2 \beta) \tilde{\phi}^{\rm B}
     - \tilde{\sigma}^{\rm B} [4a \xi \varphi^{2}_{1} - \mu D\phi +
     \Delta\phi D] \psi_{\rm D} \nonumber \\
     \!\! & & \!\! + D\phi \tilde{\lambda}^{\rm B} (D + \rho) \psi_{\rm D} 
     - [(D + \rho) \psi_{\rm D}] (\delta - 2 \beta) \tilde{\phi}^{\rm B}, 
     \nonumber \\
V_{m} \!\! & \equiv & \!\!- \psi_{\rm G} (\overline{\delta} + 4
     \overline{\beta}) \Psi^{\rm B}_{0} - \psi_{\rm H} (\overline{\delta} 
     + 4 \overline{\beta}) \overline{\Psi}^{\rm B}_{4} - \psi_{\rm E}
     (\overline{\delta} + 4 \overline{\beta}) \tilde{\sigma}^{\rm B} + 
     \psi_{\rm F} (\overline{\delta} + 4 \overline{\beta}) 
     \tilde{\lambda}^{\rm B} - \Delta\phi \psi_{\rm D} (\overline{\delta} 
     + 4\overline{\beta}) \tilde{\sigma}^{\rm B} \nonumber \\
     \!\! & & \!\! + D\phi \psi_{\rm D} (\overline{\delta} + 4
     \overline{\beta}) \tilde{\lambda}^{\rm B} - \psi_{\rm D}
     (\overline{\delta} + 4 \overline{\beta}) (\delta - 2 \beta)
     \tilde{\phi}^{\rm B}, \nonumber \\
V_{\overline{m}} \!\! & \equiv & \!\! \Psi^{\rm B}_{0} (\delta + 4 \beta)
     \psi_{\rm G} + \overline{\Psi}^{\rm B}_{4} (\delta + 4 \beta)
     \psi_{\rm H} + \tilde{\sigma} (\delta + 4 \beta) \psi_{\rm E} -
     \tilde{\lambda}^{\rm B} (\delta + 4 \beta) \psi_{\rm F} + \Delta\phi
     \tilde{\sigma}^{\rm B} (\delta + 4 \beta) \psi_{\rm D} \nonumber \\
     \!\! & & \!\! - D\phi \tilde{\lambda}^{\rm B} (\delta + 4 \beta)
     \psi_{\rm D} + (\delta + 4 \beta) \psi_{\rm D} (\delta - 2 \beta)
     \tilde{\phi}^{\rm B}.
\end{eqnarray}

Therefore, considering that in the Newman-Penrose formalism $\partial_{\mu} 
l^{\mu} = - 2 \rho$, $\partial_{\mu} n^{\mu} = 2 \mu - 2 \gamma$, 
$\partial_{\mu} m^{\mu} = 2\beta$, the continuity equation (46) can be 
rewritten in the following form:
\begin{equation}
     \partial_{\mu} (V_{l} l^{\mu} + V_{n} n^{\mu} + V_{m} m^{\nu} +
     V_{\overline{m}} \overline{m}^{\mu}) = (D - 2\rho) V_{l} + (\Delta +
     2 \mu - 2 \gamma) V_{n} + (\delta + 2 \beta) V_{m} +
     (\overline{\delta} + 2 \overline{\beta}) V_{\overline{m}} = 0.
\end{equation}

However, there is an immediate reduction in the terms involving $V_{m}$
and $V_{\overline{m}}$ in Eq.\ (57). Considering that all components of
$(\Psi^{\rm B})$ have spin weight 2 (see Eqs.\ (36), (B8), and (B9)), we
can obtain an equation analogous to Eq.\ (52):
\begin{equation}
    (\delta - 2 \beta) (\overline{\delta} + 4 \overline{\beta}) (\Psi^{\rm
    B}) = - \frac{L^{2}}{2R^{2}} (\Psi^{\rm B});
\end{equation}
furthermore, from the explicit forms of $V_{m}$ and $V_{\overline{m}}$
in Eqs.\ (56), $(\delta + 2 \beta) V_{m} + (\overline{\delta} + 2
\overline{\beta}) V_{\overline{m}}$ in Eq.\ (57) contains terms of the
form $ - (\delta + 2 \beta)[\psi_{\rm I} (\overline{\delta} + 4 
\overline{\beta}) \Psi^{\rm B}] + (\overline{\delta} + 4 
\overline{\beta}) [\Psi^{\rm B} (\delta + 4 \beta) \psi_{\rm I}]$, which,
using Eqs.\ (52) and (58), vanish:
\begin{eqnarray}
  \!\! & & \!\! - (\delta + 2 \beta) [\psi_{\rm I} (\overline{\delta} + 4
\overline{\beta}) \Psi^{\rm B}] + ({\overline{\delta}} + 4
\overline{\beta}) [\Psi^{\rm B} (\delta + 4 \beta) \psi_{\rm I}] = - 
\psi_{\rm I} (\delta - 2 \beta) (\overline{\delta} + 4 \overline{\beta}) 
\Psi^{\rm B} \nonumber \\
  \!\! & & \!\! + \Psi^{\rm B} (\overline{\delta} - 2 \overline{\beta})
(\delta + 4 \beta) \psi_{\rm I} = - \psi_{\rm I} \left[
- \frac{L^{2}}{2R^{2}} \Psi^{\rm B} \right] + \Psi^{\rm B} \left[
- \frac{L^{2}}{2R^{2}} \psi_{\rm I} \right] = 0. \nonumber
\end{eqnarray}

In this manner $(\delta + 2 \beta) V_{m} + (\overline{\delta} + 2
\overline{\beta}) V_{\overline{m}} = 0$, is satisfied identically, and
Eq.\ (57) reduces to:
\begin{equation}
     (D - 2 \rho) V_{l} + (\Delta + 2 \mu - 2 \rho) V_{n} = 0.
\end{equation}
Thus, the whole physical information about our conserved quantities is
contained in $V_{l}$ and $V_{n}$. Furthermore, direct substitutions of the 
separable solutions for the potentials (Eq.\ (51)), and field variation 
(Eqs.\ (B8) and (B9)) into the expressions for the bilinear products 
$V_{l}$ and $V_{n}$ given in Eqs.\ (56), lead to a splitting of such 
products in terms of the form $e^{0}$ and $e^{-2i\omega t}$:
\begin{eqnarray}
      V_{n} \!\! & = & \!\! [V_{n}^{+} + \frac{i\omega}{\chi^{2}}
G^{+}] \, {_{-2}}Y_{lm} \,  \overline{{_{-2}}Y_{lm}} + e^{-2i\omega t} 
[V_{n}^{-} + \frac{i\omega}{\chi^{2}}G^{-}] \, {_{-2}}Y_{lm} \, 
{_{2}}Y_{lm}, \nonumber \\
      V_{l} \!\! & = & \!\! [V_{l}^{+} + \frac{i\omega}{2} G^{+}] \,
{_{-2}}Y_{lm} \, \overline{{_{-2}}Y_{lm}} + e^{-2i\omega t} [V_{l}^{-} 
- \frac{i\omega}{2}G^{-}] \, {_{-2}}Y_{lm} \, {_{2}}Y_{lm},
\end{eqnarray}
where
\begin{eqnarray}
V_{n}^{\pm} \!\! & \equiv & \!\! \Psi^{{\rm B}\pm}_{0} [\psi_{e} - R^{3}
    \partial_{r} (R^{-3} \psi_{g})] + \psi_{h} [R^{-1} \partial_{r} (R
    \overline{\Psi}_{4}^{{\rm B}\pm})  + 8 \xi \varphi^{2}_{1}
    \tilde{\lambda}^{{\rm B}\pm}]  \nonumber \\
    \!\! & & \!\! + \psi_{e} R^{-2} \xi^{-1/2} \partial_{r} (R^{2}
    \xi^{1/2} \tilde{\sigma}^{{\rm B}\pm}) + a \psi_{f} [\Delta\phi 
    \tilde{\sigma}^{{\rm B}\pm} + \tilde{\phi}^{{\rm B}\pm}] + 
    \tilde{\lambda}^{{\rm B}\pm} R^{2} \partial_{r} (R^{2} \psi_{f}) 
    \nonumber \\
    \!\! & & \!\! - \tilde{\sigma}^{{\rm B}\pm} [\Delta\phi R
    \partial_{r} (R^{-1} \psi_{d}) + 4 a \xi \varphi^{2}_{1} \psi_{d}] + R
    [D\phi \tilde{\lambda}^{{\rm B}\pm} - \tilde{\phi}^{{\rm B}\pm}] 
    \partial_{r} (R^{-1} \psi_{d}), \nonumber \\
V_{l}^{\pm} \!\! & \equiv & \!\! \overline{\Psi}_{4}^{{\rm B}\pm} \left[
    \psi_{f} + \frac{1}{2} \chi^{-2} R^{3} \partial_{r} (R^{-3}
    \chi^{2} \psi_{h}) \right] - \psi_{g} \left[ \frac{1}{2} \chi^{-2}
    R^{-1} \partial_{r} (R\chi^{4} \Psi^{{\rm B}\pm}_{0}) + 8 \xi
    \varphi^{2}_{1} \tilde{\sigma}^{{\rm B}\pm} \right] \nonumber \\
    \!\!& & \!\!  + \frac{1}{2} \psi_{f} \chi^{4} R^{-2} \xi^{-1/2}
    \partial_{r}(\xi^{1/2} R^{2} \chi^{-2} \tilde{\lambda}^{{\rm B}\pm})
    + a \psi_{e} [\tilde{\phi}^{{\rm B}\pm} - \Delta\phi
    \tilde{\lambda}^{{\rm B}\pm}] + \frac{1}{2} 
    \tilde{\sigma}^{{\rm B}\pm} \chi^{4} R^{-2} \partial_{r} (R^{2} 
    \chi^{-2} \psi_{e}) \nonumber \\
    \!\! & & \!\! + \Delta\phi \psi_{d} \left[ - \frac{1}{2} \chi^{2} 
    R^{-1} \partial_{r} (R \tilde{\sigma}^{{\rm B}\pm}) + \frac{2a\xi
    \varphi^{2}_{1}}{D\phi} \tilde{\sigma}^{{\rm B}\pm} \right]
    \nonumber \\
    \!\! & & \!\! + D\phi \psi_{d} \left[ \frac{1}{2} \chi^{4} R^{-1}
    \partial_{r} (R \chi^{-2} \tilde{\lambda}^{{\rm B}\pm}) + \frac{2a\xi
    \varphi^{2}_{1}}{D\phi} \tilde{\lambda}^{{\rm B}\pm} \right]
    - \frac{1}{2} \chi^{2} R^{-3} \psi_{d} \partial_{r} (R^{3} 
    \tilde{\phi}^{{\rm B}\pm}),
\end{eqnarray}
and
\begin{eqnarray}
G^{+} \!\! & \equiv & \!\! \psi_{g} \Psi^{{\rm B}+}_{0} + \psi_{h}
    \overline{\Psi}^{{\rm B}+}_{4} + \psi_{e} \tilde{\sigma}^{{\rm B}+}
    - \psi_{f} \tilde{\lambda}^{{\rm B}+} + \psi_{d} [\Delta\phi
    \tilde{\sigma}^{{\rm B}+} - D\phi \tilde{\lambda}^{{\rm B}+} +
    \tilde{\phi}^{{\rm B}+}], \\
G^{-} \!\! & \equiv & \!\! \psi_{g} \Psi^{{\rm B}-}_{0} - \psi_{h}
    \overline{\Psi}^{{\rm B}-}_{4} - \psi_{e} \tilde{\sigma}^{{\rm B}-}
    - \psi_{f} \tilde{\lambda}^{{\rm B}-} + \psi_{d} [\Delta\phi
    \tilde{\sigma}^{{\rm B}-} - D\phi \tilde{\lambda}^{{\rm B}-} +
    \tilde{\phi}^{{\rm B}-}] = \frac{L^{2}l(l+1)}{8R^{4}}
    \psi^{2}_{d}, \nonumber
\end{eqnarray}
are only functions of $r$, and the relations (B10) have been used for
reducing $G^{-}$. Since the components $(\Psi^{\rm B})^{-}$ (see Eqs.\
(B8) and (B9)) are directly proportional to the potentials, $V_{n}^{-}$
and $V_{l}^{-}$ in Eqs.\ (61) have remarkable reductions (unlike
$V^{+}_{n}$ and $V^{+}_{l}$):
\begin{eqnarray}
V^{-}_{n} \!\! & = & \!\! - \frac{L^{2}l(l+1)}{8R^{4}} \left[ -2
    \psi_{g} \psi_{h} \partial_{r} \ln R^{3} + R \psi_{d} \partial_{r}
    \left( \frac{\psi_{d}}{R} \right) \right], \nonumber \\
V^{-}_{l} \!\! & = & \!\! - \frac{L^{2}l(l+1) \chi^{2}}{16R^{4}} \left[ 2
    \psi_{g} \psi_{h} \partial_{r} \ln R^{3} + R \psi_{d} \partial_{r}
    \left( \frac{\psi_{d}}{R} \right) \right],
\end{eqnarray}
therefore, from Eqs.\ (62) and (63) is very easy to show that:
\begin{eqnarray}
& & V_{n}^{-} + 2 \chi^{-2} V_{l}^{-} + R^{-2} \partial_{r} (R^{2}G^{-}) =
        0, \nonumber \\
& & V_{l}^{-} - \frac{\chi^{2}}{2} V_{n}^{-} = - \frac{L^{2}l(l+1)
    \chi^{2}}{4R^{4}} (\partial_{r} \ln R^{3}) \psi_{g} \psi_{h},
\end{eqnarray}
which will be useful below. \\[2em]

\noindent {\uno 6.3 Conserved quantities}
\vspace{.5cm}

Substituting expressions (60) into Eq.\ (59), using the explicit form for
$D$, $\Delta$, $\rho$, $\mu$, and $\gamma$ we obtain, after some 
simplification and suitably grouping, that:
\begin{eqnarray}
& & \frac{1}{R^{2}} \partial_{r} R^{2} \left[ V^{+}_{l} -
    \frac{\chi^{2}}{2} V^{+}_{n} \right] {_{-2}}Y_{lm} \,
    \overline{{_{-2}}Y_{lm}} + \frac{e^{-2i\omega t}}{R^{2}}
    \partial_{r} R^{2} \left[ V^{-}_{l} - \frac{\chi^{2}}{2} V^{-}_{n} 
    \right] {_{-2}}Y_{lm} \, {_{2}}Y_{lm} \nonumber \\
& & - i \omega e^{-2i\omega t} \left[ 2 \frac{V_{l}^{-}}{\chi^{2}} 
    + V^{-}_{n} + R^{-2} \partial_{r} (R^{2} G^{-}) \right] {_{-2}}Y_{lm}
    \, {_{2}}Y_{lm} = 0 ,
\end{eqnarray}
the last term vanishes in according to the first of Eqs.\ (64), thus Eq.\
(65) reduces to:
\begin{equation}
\partial_{r} R^{2} \left[ V^{+}_{l} - \frac{\chi^{2}}{2} V^{+}_{n} \right]
    {_{-2}}Y_{lm} \, \overline{{_{-2}}Y_{lm}} + e^{-2i\omega t}
    \partial_{r} R^{2} \left[ V^{-}_{l} - \frac{\chi^{2}}{2}
    V^{-}_{n} \right] {_{-2}}Y_{lm} \, {_{2}}Y_{lm} = 0,
\end{equation}
which implies (using the linear independence of terms of the form
$e^{i\omega t}$ and $e^{-i\omega t}$) that there exist two conserved
quantities, which we denote by $K^{(\pm)}$:
\begin{equation}
R^{2} \left[ V^{(\pm)}_{l} - \frac{\chi^{2}}{2} V^{(\pm)}_{n} \right]
    \equiv K^{(\pm)}.
\end{equation}
Although $K^{+}$ has a complicated form in terms of the potentials,
$K^{-}$ has a remarkably simple form, in accordance with the last
expression in Eq.\ (64):
\begin{equation}
K^{-} \equiv R^{2} \left[ V^{-}_{l} - \frac{\chi^{2}}{2} V^{-}_{n} \right]
    = - \frac{L^{2}l(l+1)}{4} \frac{\chi^{2}(\partial_{r} \ln
    R^{3})}{R^{2}} \psi_{g} \psi_{h}. 
\end{equation}
Note that, since $(\Psi^{\rm B})^{+}$ depends on $(\overline{\psi}_{i})$,
$K^{+}$ depends on $(\psi_{i})$ and $(\overline{\psi_{i}})$, whereas
$K^{-}$ directly on the potentials without involving its complex 
conjugates.

The existence of these two conserved quantities deserves some important 
comments. First: although the equations used for obtaining such quantities 
are not Hermitian ones (for which the constancy of the Wronskian yields 
traditionally conserved quantities), one can obtain, without any 
restrictions and full generality, conserved quantities, provided that the 
original system of equations and its adjoint system to be used. Second: as
we have seen, if the potentials have a time dependence of the form 
$e^{-i\omega t}$, the field perturbations appearing in the decoupled
system contain terms proportional to $e^{-i\omega t}$ and $e^{i\omega t}$ 
(in the classical cases, unlike the present case involving string fields, 
only terms proportional to $e^{i\omega t}$ are present \cite{18}), which 
lead finally to two conserved quantities. In the classical cases, only a 
conserved quantity analogous to the present $K^{+}$ is obtained. In fact, 
the bilinear terms depending on $\Psi^{{\rm B}+}_{0}$ and $\psi_{g}$ in 
the expression for $K^{+}$ (see the explicit forms for $V^{+}_{n}$ and 
$V^{+}_{l}$ in Eqs.\ (61)), yield a conservation relation for the energy
of gravitational perturbations in the classical Schwarzschild black hole 
(and something similar for electromagnetic perturbations) \cite{18}. In
this manner, it is possible that $K^{+}$ has the same physical meaning 
for the present string black hole: the conservation of the energy for the 
coupled field perturbations. However, this question will require a long 
asymptotic analysis and, will be studied in a subsequent work. On the 
other hand, $K^{-}$ is a novel conserved quantity apparently without 
classical analogous; it is also an open question to investigate its 
physical meaning. \\[2em]

\noindent{\uno 6.4 Differential identities}
\vspace{.5cm}

As mentioned, $(\Psi^{\rm B})$ in the decoupled system can be expressed 
essentially in the form $(\Psi^{\rm B}) = (\Psi^{\rm B})^{+}  \, 
\overline{{_{-2}}Y_{lm}} \, e^{i\omega t} + (\Psi^{\rm B})^{-} \, 
{_{2}}Y_{lm} \, e^{-i\omega t}$. Thus, the decoupled system ${\cal O} 
(\Psi^{\rm B}) = 0$, can be reduced (again, using the linear independence 
of the terms of the form $e^{i\omega t}$ and $e^{-i\omega t}$) to ${\cal
O} (\Psi^{\rm B})^{+} = 0$, and ${\cal O} (\Psi^{\rm B})^{-} = 0$. The 
adjoint system for the potentials is the same, coming from both above 
equations: ${\cal O}^{\dag} (\psi) = 0$. In this manner, the two conserved 
quantities constructed in Section 6.3, can be obtained separately: $K^{+}$
will become from the equation $(\psi) {\cal O} (\Psi^{\rm B})^{+} - {\cal 
O}^{\dag} (\psi) (\Psi^{\rm B})^{+} = \nabla_{\mu} J_{+}^{\mu}$ and 
$K^{-}$ will from the equation $(\psi) {\cal O} (\Psi^{\rm B})^{-} - 
{\cal O}^{\dag} (\psi) (\Psi^{\rm B})^{-} = \nabla_{\mu} J_{-}^{\mu}$. 
However, ${\cal O} (\Psi^{\rm B})^{-} = 0$ is essentially the same 
equations for the potentials ${\cal O}^{\dag}(\psi) = 0$ (remembering that
the components of $(\Psi^{\rm B})^{-}$ are directly proportional to 
$(\psi)$). In fact, after separation of variables, the first row of 
equations ${\cal O} (\Psi^{\rm B})^{-} = 0$ corresponds to the second 
equation for the  potentials (which means, the second row of ${\cal 
O}^{\dag} (\psi) = 0$), satisfying the following differential identities 
between components of the operators ${\cal O}$ and ${\cal O}^{\dag}$: 
$R^{4} {\cal O}_{11} \frac{1}{R^{4}} = {\cal O}^{\dag}_{22}$, and 
$\frac{1}{8 \varphi_{1}^{2}} ({\cal O}_{13} - F_{1} \Delta\phi) \xi^{-1} 
= {\cal O}^{\dag}_{42}$. Similarly, the second of those equations, 
corresponds to the first equation for the potentials satisfying the 
relations $R^{4} {\cal O}_{22} \frac{1}{R^{4}} = {\cal O}^{\dag}_{11}$, 
and $- \frac{1}{8 \varphi_{1}^{2}} [ {\cal O}_{24} + \frac{\chi^{4}}{4} 
F_{1} D\phi ] \xi^{-1} = {\cal O}^{\dag}_{31}$. The third and fourth of
the decoupled equations correspond to the following combinations of the 
equations for the potentials: (fourth one) + $D\phi$ (fifth one) and, 
(third one) -- $\Delta\phi$ (fourth one) respectively. In these cases, the 
following differential identities are satisfied:
\begin{eqnarray}
2 Q^{2} \xi {\cal O}_{31} \frac{1}{R^{4}} \!\! & = & \!\! 
     {\cal O}^{\dag}_{24} + \frac{\chi^{4}}{4} F_{1} D\phi, \nonumber \\
- \frac{a}{2} F_{1} \!\! & = & \!\! {\cal O}^{\dag}_{34} + D\phi {\cal  
     O}^{\dag}_{35}, \nonumber \\
\xi ({\cal O}_{33} - {\cal O}_{35} \Delta\phi) \xi^{-1} \!\! & = & \!\! 
     - ({\cal O}^{\dag}_{44} + D\phi {\cal O}^{\dag}_{45}), \nonumber \\  
Q^{2} \xi {\cal O}_{35} \frac{1}{R^{4}} \!\! & = & \!\! 
     {\cal O}^{\dag}_{54} + D\phi {\cal O}^{\dag}_{55},  \\
\noalign{\hbox{and}}
- 2 Q^{2} \xi {\cal O}_{42} \frac{1}{R^{4}} \!\! & = & \!\! 
     {\cal O}^{\dag}_{13} - F_{1} \Delta\phi, \nonumber \\
\frac{a \chi^{4}}{8} F_{1} \!\! & = & \!\! {\cal O}^{\dag}_{43} - \Delta 
     \phi {\cal O}^{\dag}_{45}, \nonumber \\
\xi ({\cal O}_{44} + {\cal O}_{45} D\phi) \xi^{-1} \!\! & = & \!\! -
     ({\cal O}^{\dag}_{33} - \Delta\phi {\cal O}^{\dag}_{35}), \nonumber
     \\
- Q^{2} \xi {\cal O}_{45} \frac{1}{R^{4}} \!\! & = & \!\! 
     {\cal O}^{\dag}_{53} - \Delta\phi {\cal O}^{\dag}_{55},
\end{eqnarray}
respectively. Finally, the fifth of the decoupled equations corresponds to
the fifth of the equations for the potentials, and the corresponding
differential identities are:
\begin{eqnarray}
R^{4} {\cal O}_{55} \frac{1}{R^{4}} \!\! & = & \!\! {\cal O}^{\dag}_{55},
     \nonumber \\
\frac{1}{4\varphi^{2}_{1}} ({\cal O}_{54} + {\cal O}_{55} D\phi) \xi^{-1}
     \!\! & = & \!\! - {\cal O}^{\dag}_{35}, \nonumber \\
\frac{1}{4\varphi^{2}_{1}} ({\cal O}_{53} - {\cal O}_{55} \Delta\phi)
     \xi^{-1} \!\! & = & \!\! {\cal O}^{\dag}_{45}.
\end{eqnarray}

What do such differential identities mean? The answer is that they map
solutions of the equations for the (radial parts) of the potentials into
solutions for the (radial parts) of the field variations appearing in the
decoupled set of equations, and conversely. 

As we have demonstrated, if
\begin{equation}
   (\psi) (r) = \pmatrix{ \psi_{g} \cr \psi_{h} \cr \psi_{e} \cr \psi_{f}
   \cr \psi_{d} \cr},
\end{equation}
is the radial part of a solution of the form $(\psi) = (\psi) (r) \,
{_{-2}Y_{lm}} \, e^{-i\omega t}$ admitted by ${\cal O}^{\dag} (\psi) = 0$,
then
\begin{equation}
   (\Psi^{\rm B})^{-} (r) = \pmatrix{ \frac{1}{R^{4}} \psi_{h} \cr
\frac{1}{R^{4}} \psi_{g} \cr - \frac{1}{2Q^{2}\xi} \psi_{f} \cr
\frac{1}{2Q^{2}\xi} \psi_{e} \cr \frac{1}{2} \left( \frac{\psi_{d}}{R^{4}}
+ \frac{1}{Q^{2}\xi} (D\phi \psi_{e} + \Delta\phi \psi_{f}) \right) \cr},
\end{equation}
is the radial part of a solution of the form $(\Psi^{\rm B}) = (\Psi^{\rm
B})^{-} (r) \, {_{2}}Y_{lm} \, e^{-i\omega t}$ for the decoupled system
${\cal O} (\Psi^{\rm B}) = 0$. If in the preceding expression for
$(\Psi^{\rm B})$, $\omega$ is replaced by $- \omega$, then $(\Psi^{\rm
B}) = (\overline{\Psi^{\rm B}})^{-} \, {_{2}}Y_{lm} \, e^{i\omega t}$
satisfies ${\cal O} (\Psi^{\rm B}) = 0$ with
\begin{equation}
   (\overline{\Psi^{\rm B}})^{-} (r) = \pmatrix{ \frac{1}{R^{4}}
\overline{\psi}_{h} \cr \frac{1}{R^{4}} \overline{\psi}_{g} \cr 
- \frac{1}{2Q^{2}\xi} \overline{\psi}_{f} \cr \frac{1}{2Q^{2}\xi}
\overline{\psi}_{e} \cr \frac{1}{2} \left(
\frac{\overline{\psi}_{d}}{R^{4}} + \frac{1}{Q^{2}\xi} (D\phi
\overline{\psi}_{e} + \Delta\phi \overline{\psi}_{f}) \right) \cr}.
\end{equation}

On the other hand, $(\Psi^{\rm B})^{+}$ in Eq.\ (B8) and (B9) is also the
radial part of a solution of the form $e^{i\omega t}$ for the decoupled
system. Thus, $(\Psi^{\rm B})^{+} = C (\overline{\Psi^{\rm B}})^{-}$,
being $C$ a constant. This relation of proportionality would lead to
differential identities analogous to the Teukolsky-Starobinsky identities
found in the study of classical black holes \cite{21}. However, this
subject will be extended in a subsequent work. \\[2em]

\begin{center}
{\uno VII. CONCLUDING REMARKS}
\end{center}
\vspace{1em}

We summarize some questions that remain open and will be the subject of
forthcoming
works.

First: although string black holes are considered as classical black holes
plus Planck-scale corrections, they are not actually {\it authentic
quantum black holes}. Hence, for example, the thermodynamics properties
argued in Refs.\ \cite{2,3} are limited in this sense; a proper 
quantization will give a more complete and satisfactory description of 
such objects (see the paragraph before final comments of Ref.\ \cite{5}). 
The idea is, of course, that the symplectic structure constructed in the 
present work, to be the starting point for such a proper (canonical) 
quantization, which will give us a consistent quantum extension of string 
black holes.

Second: as mentioned, the physical meaning of the conserved quantities
obtained in the present work, remains to be worked out. This subject will
include the calculation of physical quantities such as scattering
amplitudes, reflection and transmission coefficients, etc. The
differential identities established here, will be useful in this task; 
they will permit to relate the outcoming flux of energy to the incoming 
flux of energy for the coupled field perturbations \cite{21}.

Third: the results established in Sec.\ II can be considered in the formal
context of differential equations. The possible applications of these very 
general results in other cases (and other areas of physics) are open 
questions. 

Finally, beyond the specific application presented in this work, adjoint
operators scheme gives a new approach for covariant canonical
quantization\cite{22}, which represents a subject of permanent and wide
interest in
physics. The possible implications by using this approach in this matter
is also a problem for the future.  \\[2em]

\begin{center}
{\uno ACKNOWLEDGMENTS}
\end{center}
\vspace{1em}

This work was supported by CONACYT and the Sistema Nacional de
Investigadores (M\'exico). \\[2em]

\begin{center}
{\uno Appendix A: Gauge invariant perturbations}
\end{center}
\vspace{1em}
\renewcommand{\theequation}{A\arabic{equation}}
\setcounter{equation}{0}
\vspace{1em}

In order to construct quantities with invariance properties similar those
of $\tilde{\sigma}^{\rm B}$, which are useful in our approach, we follow 
Eqs.\ (15) and (16), and we find the following expression for the 
variations of the vanishing background Newman-Penrose quantities:
\begin{eqnarray}
\kappa^{\rm B} \!\! & \equiv & \!\! - (l^{\mu} l^{\nu} \nabla_{\nu}
     m_{\mu})^{\rm B} = l^{\mu} l^{\nu} m_{\gamma}
     (\Gamma^{\gamma}_{\mu\nu})^{\rm B} - (D - \rho) (l^{\mu} m^{\rm
     B}_{\mu}),   \nonumber \\
\overline{\pi}^{\rm B} \!\! & \equiv & \!\! - (m^{\mu} l^{\nu}
     \nabla_{\nu} n_{\mu})^{\rm B} = m^{\mu} l^{\nu} n_{\gamma}
     (\Gamma^{\gamma}_{\mu\nu})^{\rm B} - D (m^{\mu} n^{\rm B}_{\mu}) + 
     \mu (l^{\mu} m^{\rm B}_{\mu}), \nonumber \\
\overline{\lambda}^{\rm B} \!\! & \equiv & \!\! - (m^{\mu} m^{\nu}
     \nabla_{\nu} n_{\mu})^{\rm B} = m^{\mu} m^{\nu} n_{\gamma}
     (\Gamma^{\gamma}_{\mu\nu})^{\rm B} + \mu m^{\mu} m^{\nu} h_{\mu\nu} -
     (\delta - 2 \beta) (m^{\mu} n^{\rm B}_{\mu}), \nonumber \\
\overline{\nu}^{\rm B} \!\! & \equiv & \!\! - (m^{\mu} n^{\nu}
     \nabla_{\nu} n_{\mu})^{\rm B} = m^{\mu} n^{\nu} n_{\gamma}
     (\Gamma^{\gamma}_{\mu\nu})^{\rm B} + \mu m^{\mu} n^{\nu} h_{\mu\nu} -
     (\Delta + 2 \gamma + \mu) (m^{\mu} n^{\rm B}_{\mu}), \nonumber \\
\tau^{\rm B} \!\! & \equiv & \!\! - (l^{\mu} n^{\nu} \nabla_{\nu}
     m_{\mu})^{\rm B} = l^{\mu} n^{\nu} m_{\gamma}
     (\Gamma^{\gamma}_{\mu\nu})^{\rm B} - (\Delta - 2 \gamma) (l^{\mu}
     m^{\rm B}_{\mu}) + \rho (n^{\mu} m^{\rm B}_{\mu}), \nonumber \\
\overline{\varphi}^{\rm B}_{2} \!\! & \equiv & \!\!  (m^{\mu} n^{\nu}
     F_{\mu\nu})^{\rm B} = m^{\mu} n^{\nu} F_{\mu\nu}^{\rm B} - 2
     \varphi_{1} (n^{\mu} m^{\rm B}_{\mu}), \nonumber \\
(\delta \varphi_{1})^{\rm B} \!\! & = & \!\! \delta \varphi^{\rm B}_{1}
     - 2 \rho \varphi_{1} m^{\mu} n^{\rm B}_{\mu} + 2 \mu \varphi_{1}
     m^{\mu} l^{\rm B}_{\mu}, \nonumber \\
(\delta \overline{\varphi}_{1})^{\rm B} \!\! & = & \!\! \delta
     \overline{\varphi}^{\rm B}_{1} + 2 \rho \varphi_{1} m^{\mu}
     n^{\rm B}_{\mu} - 2 \mu \varphi_{1} m^{\mu} l^{\rm B}_{\mu},
     \nonumber \\
(\delta \phi)^{\rm B} \!\! & = & \!\! \delta \phi^{\rm B} - D \phi
     (m^{\mu} n^{\rm B}_{\mu}) - \Delta \phi (m^{\mu} l^{\rm B}_{\mu}),
\end{eqnarray}
where $m^{\mu} n^{\rm B}_{\mu}$, $n^{\mu} m^{\rm B}_{\mu}$, $m^{\mu}
l^{\rm B}_{\mu}$, and $l^{\mu} m^{\rm B}_{\mu}$ are dependent on the 
perturbed tetrad gauge freedom and Eqs.\ (6)-(8) for the background 
quantities have been considered. Note that
\begin{equation}
     2 \varphi^{\rm B}_{1} = (l^{\mu} n^{\nu} + \overline{m}^{\mu}
     m^{\nu}) F^{\rm B}_{\mu\nu} - 2 \varphi_{1} [m_{\mu} 
     (\overline{m}^{\mu})^{\rm B} + \overline{m}_{\mu} (m^{\mu})^{\rm B}]
     = (l^{\mu} n^{\nu} + \overline{m}^{\mu} m^{\nu}) F^{\rm B}_{\mu\nu} 
     + 2 \varphi_{1} \overline{m}^{\mu} m^{\nu} h_{\mu\nu},
\end{equation}
which means that $\varphi^{\rm B}_{1} = \varphi^{\rm B}_{1} (h_{\mu\nu},
b_{\mu})$, is defined completely in terms of $h_{\mu\nu}$ and $b_{\mu}$,
and independent on the perturbed tetrad gauge freedom. Thus, from Eqs.\ 
(A1) and (A2) we can find easily the following quantities, independent on
both, perturbed tetrad gauge freedom and gauge transformations of the
vector potential variations:
\begin{eqnarray}
\tilde{\sigma}^{\rm B} \!\! & \equiv & \!\! \sigma^{\rm B} + (\delta - 2 
     \beta) \frac{\varphi^{\rm B}_{0}}{2 \varphi_{1}}, \nonumber \\
\tilde{\kappa}^{\rm B} \!\! & \equiv & \!\! \kappa^{\rm B} + (D - \rho)
     \frac{\varphi^{\rm B}_{0}}{2 \varphi_{1}}, \nonumber \\
\tilde{\pi}^{\rm B} \!\! & \equiv & \!\! \overline{\pi}^{\rm B} + D \left(
     \frac{\overline{\varphi}^{\rm B}_{2}}{2 \varphi_{1}} \right) - \mu
     \frac{\varphi^{\rm B}_{0}}{2 \varphi_{1}}, \nonumber \\
\tilde{\lambda}^{\rm B} \!\! & \equiv & \!\! \overline{\lambda}^{\rm B}
     + (\delta - 2 \beta) \frac{\overline{\varphi}^{\rm B}_{2}}{2 
     \varphi_{1}}, \nonumber \\
\tilde{\nu}^{\rm B} \!\! & \equiv & \!\! \overline{\nu}^{\rm B} + (\Delta
     + 2 \gamma + \mu) \frac{\overline{\varphi}^{\rm B}_{2}}{2
     \varphi_{1}}, \nonumber \\
\tilde{\tau}^{\rm B} \!\! & \equiv & \!\! \tau^{\rm B} + (\Delta - 2
     \gamma) \frac{\varphi^{\rm B}_{0}}{2 \varphi_{1}} + \rho
     \frac{\overline{\varphi}^{\rm B}_{2}}{2 \varphi_{1}}, \nonumber \\
\hat{\varphi}^{\rm B}_{1} \!\! & \equiv & \!\! (\delta
     \varphi_{1})^{\rm B} + \mu \varphi^{\rm B}_{0} + \rho
     \overline{\varphi}^{\rm B}_{2}, \qquad  \check{\varphi}^{\rm B}_{1}
     \equiv (\delta \overline{\varphi}_{1})^{\rm B} - \mu \varphi^{\rm
     B}_{0} - \rho \overline{\varphi}^{\rm B}_{2}, \nonumber \\
\tilde{\phi}^{\rm B} \!\! & \equiv & \!\! (\delta \phi)^{\rm B} - \Delta
     \phi \frac{\varphi^{\rm B}_{0}}{2 \varphi_{1}} + D \phi
     \frac{\overline{\varphi}^{\rm B}_{2}}{2 \varphi_{1}}.
\end{eqnarray}
The variations of the Weyl scalars $\Psi^{\rm B}_{0}$, and
$\overline{\Psi}^{\rm B}_{4}$ turn out to be directly, independent on the
perturbed tetrad gauge freedom, similar to the perturbed quantity in Eq.\
(A2). Finally, we can find the following gauge invariant quantities,
related to the Weyl scalar variations and electromagnetic field 
variations:
\begin{eqnarray}
\tilde{\Psi}^{\rm B}_{3} \!\! & \equiv & \!\! \overline{\Psi}^{\rm B}_{3}
     + 3 \Psi_{2} \left( \frac{\overline{\varphi}^{\rm B}_{2}}{2
     \varphi_{1}} \right), \nonumber \\
\tilde{\Psi}^{\rm B}_{1} \!\! & \equiv & \!\! \Psi^{\rm B}_{1} - 3
     \Psi_{2} \left( \frac{\varphi^{\rm B}_{0}}{2 \varphi_{1}} \right).
\end{eqnarray}

In this manner, the field quantities in Eqs.\ (A3), and (A4) (and
${\Psi}^{\rm B}_{0}$, and $\overline{\Psi}^{\rm B}_{4}$ in according to
first and fifth of Eqs.\ (A21)) are determined completely in terms of
$h_{\mu\nu}$, $b_{\mu}$, and $\phi^{\rm B}$.

With the purpose of finding the equations governing the gauge invariant
variations, let us take first-order variations of Eq.\ (A3) of Ref.\
\cite{12}, and we obtain the following equation involving no gauge 
invariance quantities:
\begin{equation}
     (\Delta - 2 \gamma + \mu - a \Delta\phi) \varphi^{\rm B}_{0} -
     (\delta \varphi_{1})^{\rm B} + 2 \varphi_{1} \tau^{\rm B} + a D\phi
     \overline{\varphi}^{\rm B}_{2} + 2 a \varphi_{1} (\delta \phi)^{\rm B}
     = \xi^{-1} m^{\mu} j_{\mu},
\end{equation}
where the background solution for the static charged black holes of Sec.\
II has been considered and a source $j_{\mu}$ for the electromagnetic
perturbations has been included \cite{12}. However, using the expressions 
(A3) we can substitute $(\delta \varphi_{1})^{\rm B}$, $\tau^{\rm B}$, 
and $(\delta \phi)^{\rm B}$ in favor of $\hat{\varphi}^{\rm B}_{1}$, 
$\tilde{\tau}^{\rm B}$, and $\tilde{\phi}^{\rm B}$, into Eq.\ (A5), and to 
obtain easily the equation
\begin{equation}
     2 \varphi_{1} \tilde{\tau}^{\rm B} + 2 a \varphi_{1} \tilde{\phi}^{\rm
     B} - \hat{\varphi}^{\rm B}_{1} = \xi^{-1} m^{\mu} j_{\mu},
\end{equation}
involving only gauge invariant quantities. Similarly, from the complex
conjugate of Eq.\ (A4) of Ref.\ \cite{12} we obtain
\begin{equation}
     - 2 \varphi_{1} \tilde{\pi}^{\rm B} + 2 a \varphi_{1} \tilde{\phi}^{\rm
     B} + \check{\varphi}^{\rm B}_{1} = \xi^{-1} m^{\mu} j_{\mu}.
\end{equation}
The remaining two Maxwell equations (A1) and (A2) of Ref.\ \cite{12}, 
require a more elaborate procedure in order to avoid the appearance of 
undesirable perturbed quantities. Before considering the variations, we 
apply $\delta$ to Eq.\ (A1) of Ref.\ \cite{12} and we obtain
\begin{equation}
     \delta (\overline{\delta} + \pi - 2 \alpha) \varphi_{0} - \delta D
     \varphi_{1} + 2 \delta (\rho \varphi_{1}) - \delta (\kappa \varphi_{2})
     - a \delta [\varphi_{0} \overline{\delta} + \overline{\varphi}_{0}
     \delta - (\varphi_{1} + \overline{\varphi}_{1}) D] \phi = 0,
\end{equation}
using the commutation relations, the second term can be expressed as
\begin{eqnarray}
     \delta D \varphi_{1} = (D - \overline{\rho} - \epsilon +
     \overline{\epsilon}) \delta \varphi_{1} + (\overline{\alpha} + \beta 
     - \overline{\pi}) D \varphi_{1} + \kappa \Delta \varphi_{1} - \sigma
     \overline{\delta} \varphi_{1}, \nonumber
\end{eqnarray}
and considering the background solution, we have from the above equation
that
\begin{equation}
     (\delta D \varphi_{1})^{\rm B} = (D - \rho) (\delta \varphi_{1})^{\rm
     B} + D \varphi_{1} (\overline{\alpha} + \beta - \overline{\pi})^{\rm B}
     + \Delta \varphi_{1} \kappa^{\rm B},
\end{equation}
thus, from Eqs.\ (A8) and (A9) and considering again the background
solution, one obtains the linearized equation
\begin{eqnarray}
\!\! & & \!\! \delta (\overline{\delta} + 2 \overline{\beta})
     \varphi^{\rm B}_{0} - (D - 3 \rho - a D\phi) (\delta
     \varphi_{1})^{\rm B} + 2 \mu \varphi_{1} \kappa^{\rm B} + 2 \rho
     \varphi_{1} \overline{\pi}^{\rm B} + a D\phi (\delta
     \overline{\varphi}_{1})^{\rm B} \nonumber \\
\!\! & & \!\! + 2 \varphi_{1} [(\delta \rho)^{\rm B} - \rho
     (\overline{\alpha} + \beta)^{\rm B}] = \delta (\xi^{-1} l^{\mu} 
     j_{\mu}),
\end{eqnarray}
however, from the Ricci identities we can find additionally the linearized
equation
\begin{equation}
     (\delta \rho)^{\rm B} - \rho (\overline{\alpha} + \beta)^{\rm B} -
     (\overline{\delta} + 4 \overline{\beta}) \sigma^{\rm B} + \Psi^{\rm
     B}_{1} - D\phi (\delta \phi)^{\rm B} - 2 \xi \varphi_{1} \varphi^{\rm
     B}_{0} = l^{\mu} m^{\nu} T_{\mu\nu},
\end{equation}
where we have included an additional source for the gravitational
perturbations, $T_{\mu\nu}$ \cite{12}, and $\Phi_{01}^{\rm B}= D\phi 
(\delta \phi)^{\rm B} + 2 \xi \varphi_{1} \varphi^{\rm B}_{0}$ (see Eqs.\
(A8) of Ref.\ \cite{12}). Therefore, we have finally, from Eqs.\ (A10),
(A11) and from direct substitutions of the relations (A3) and (A4), that
\begin{eqnarray}
\!\! & & \!\! - (D - 3 \rho - a D\phi) \hat{\varphi}^{\rm B}_{1} + a
     D \phi \check{\varphi}^{\rm B}_{1} + 2 \mu \varphi_{1}
     \tilde{\kappa}^{\rm B} + 2 \rho \varphi_{1} \tilde{\pi}^{\rm B}
     + 2 \varphi_{1} (\overline{\delta} + 4 \overline{\beta})
     \tilde{\sigma}^{\rm B} - 2 \varphi_{1} \tilde{\Psi}^{\rm B}_{1}
     \nonumber \\
\!\! & & \!\! + 2 \varphi_{1} D\phi \tilde{\phi}^{\rm B} = \delta
     (\xi^{-1} l^{\mu} j_{\mu}) - 2 \varphi_{1} l^{\mu} m^{\nu} T_{\mu\nu},
\end{eqnarray}
which involves only gauge invariant quantities. Similarly, from Eq.\
(A2) of Ref.\ \cite{12} and using the linearized equation
\begin{equation}
     - (\delta \overline{\mu})^{\rm B} - \mu (\overline{\alpha} +
     \beta)^{\rm B} + (\overline{\delta} + 4 \overline{\beta})
     \overline{\lambda}^{\rm B} + \overline{\Psi}^{\rm B}_{3} - \Delta
     \phi (\delta \phi)^{\rm B} + 2 \xi \varphi_{1} 
     \overline{\varphi}^{\rm B}_{2} = m^{\mu} m^{\nu} T_{\mu\nu},
\end{equation}
coming from the Ricci identities, we can obtain the equation
\begin{eqnarray}
     \!\! & & \!\!  (\Delta + 3 \mu - a \Delta\phi) \check{\varphi}^{\rm
     B}_{1} - a \Delta\phi \hat{\varphi}^{\rm B}_{1} + 2 \mu \varphi_{1}
     \tilde{\tau}^{\rm B} + 2 \rho \varphi_{1} \tilde{\nu}^{\rm B} - 2
     \varphi_{1} (\overline{\delta} + 4 \overline{\beta})
     \tilde{\lambda}^{\rm B} - 2 \varphi_{1} \tilde{\Psi}^{\rm B}_{3}
     \nonumber \\
     \!\! & & \!\! + 2 \varphi_{1} \Delta\phi \tilde{\phi}^{\rm B} = \delta
     (\xi^{-1} n^{\mu} j_{\mu}) - 2 \varphi_{1} n^{\mu} m^{\nu} T_{\mu\nu}.
\end{eqnarray}
In the case of the dilaton equation, we apply again $\delta$ to Eq.\ (A5) of
Ref.\ \cite{12}, before considering the variations:
\begin{equation}
     D\phi (\delta \overline{\mu}) + \overline{\mu} \delta D\phi + \delta
     [(D + \epsilon + \overline{\epsilon} - \rho) \Delta\phi] - (\delta
     \overline{\pi}) \overline{\delta} \phi - \pi \delta \overline{\delta}
     \phi + \delta [(- \overline{\delta} + \alpha - \overline{\beta} - \pi)
     \delta \phi] + \frac{a}{4} \delta (\xi F^{2}) = 0.
\end{equation}
Moreover, using the commutation relations (see Eq.\ (A9)) one finds that
\begin{eqnarray}
(\delta D \phi)^{\rm B} \!\! & = & \!\! (D - \rho) (\delta \phi)^{\rm B}
     + \Delta\phi \kappa^{\rm B} + D \phi (\overline{\alpha} + \beta -
     \overline{\pi})^{\rm B}, \nonumber \\
(\delta \Delta\phi)^{\rm B} \!\! & = & \!\! (\Delta + \mu) (\delta
     \phi)^{\rm B} - D\phi \overline{\nu}^{\rm B} + \Delta\phi (\tau
     - \overline{\alpha} - \beta)^{\rm B},
\end{eqnarray}
where the background solution has been considered. Furthermore,
\begin{equation}
     (\delta \xi F^{2})^{\rm B} = - 8 \xi \varphi_{1} [(\delta
     \varphi_{1})^{\rm B} - (\delta \overline{\varphi}_{1})^{\rm B} - 2 a
     \varphi_{1} (\delta \phi)^{\rm B}],
\end{equation}
where Eq.\ (A7) of Ref.\ \cite{12} has been used. Similarly,
\begin{equation}
     [\delta (D + \epsilon + \overline{\epsilon} - \rho) \Delta\phi]^{\rm
     B} = (D - 2 \rho) (\delta \Delta \phi)^{\rm B} + D \Delta\phi
     (\overline{\alpha} + \beta - \overline{\pi})^{\rm B} + \Delta^{2} \phi
     \kappa^{\rm B} + \Delta\phi [\delta (\epsilon +
     \overline{\epsilon})^{\rm B} - (\delta \rho)^{\rm B}].
\end{equation}
On the other hand, from the Ricci identities
\begin{equation}
     (D - \rho) (\overline{\alpha} + \beta)^{\rm B} - \delta (\epsilon +
     \overline{\epsilon})^{\rm B} + (\mu + 2 \gamma) \kappa^{\rm B} - \rho
     \overline{\pi}^{\rm B} - \Psi^{\rm B}_{1} - D\phi (\delta\phi)^{\rm
     B} - 2 \xi \varphi_{1} \varphi^{\rm B}_{0} = l^{\mu} m^{\nu}
     T_{\mu\nu}.
\end{equation}
Thus, by linearizing Eq.\ (A15), considering Eqs.\ (A16)--(A19) and direct
substitutions of $(\delta\rho)^{\rm B}$, $(\delta \overline{\mu})^{\rm
B}$, $\delta (\epsilon + \overline{\epsilon})^{\rm B}$ from Eqs.\ (A11),
(A13), and (A19), we have, after some simplification and grouping
suitably, that
\begin{eqnarray}
     \!\! & & \!\! [\mu (D - \rho) + (D - 2 \rho) (\Delta + \mu) - \delta
     (\overline{\delta} + 2 \overline{\beta}) - 3 D\phi \Delta\phi - 4
     a^{2} \xi \varphi^{2}_{1}] \tilde{\phi}^{\rm B} + [\Delta^{2} \phi +
     2 (\mu + \gamma) \Delta\phi] \tilde{\kappa}^{\rm B} \nonumber \\
     \!\! & & \!\! - (D \Delta\phi) \tilde{\pi}^{\rm B} + D\phi
     (\overline{\delta} + 4 \overline{\beta}) \tilde{\lambda}^{\rm B} -
     \Delta\phi (\overline{\delta} + 4 \overline{\beta})
     \tilde{\sigma}^{\rm B} - (D - 2 \rho) D\phi \tilde{\nu}^{\rm B} +
     (D - 2 \rho) \Delta\phi \tilde{\tau}^{\rm B} \nonumber \\
     \!\! & & \!\! + (D\phi) \tilde{\Psi}^{\rm B}_{3} + 2 a \xi \varphi_{1}
     (\check{\varphi}^{\rm B}_{1} - \hat{\varphi}^{\rm B}_{1}) =
     \frac{1}{2} \delta \phi_{s} + D \phi n^{\mu} m^{\nu} T_{\mu\nu} + 2 
     \Delta\phi l^{\mu} m^{\nu} T_{\mu\nu},
\end{eqnarray}
where $\phi_{s}$ represents a source for the dilaton field perturbations,
and the relations (A3) and (A4) have been considered. The above equation
involves, as wanted, only gauge invariant quantities.

The system of equations (A6), (A7), (A12), (A14), and (A20) comes from
the linearization of the matter field equations (A1)-(A5) of Ref.\ 
\cite{12}, considering that the background solution corresponds to 
dilatonic charged black holes. This system is completed by linearizing 
Ricci identities:
\begin{eqnarray}
\Psi^{\rm B}_{0} + (\delta - 2 \beta) \tilde{\kappa}^{\rm B} - (D - 2
     \rho) \tilde{\sigma}^{\rm B} \!\! & = & \!\! 0, \nonumber \\
(D - \rho) \tilde{\lambda}^{\rm B} - (\delta - 2 \beta)
     \tilde{\pi}^{\rm B} - \mu \tilde{\sigma}^{\rm B} \!\! & = & \!\!
     m^{\mu} m^{\nu} T_{\mu\nu}, \nonumber \\
- \tilde{\Psi}^{\rm B}_{1} - (\Delta - 4 \gamma) \tilde{\kappa}^{\rm B}
     + (D - \rho) \tilde{\tau}^{\rm B} - \rho \tilde{\pi}^{\rm B} - D\phi
     \tilde{\phi}^{\rm B} \!\! & = & \!\! l^{\mu} m^{\nu} T_{\mu\nu},
     \nonumber \\
(\delta - 2 \beta) \tilde{\tau}^{\rm B} - (\Delta - 2 \gamma + \mu)
     \tilde{\sigma}^{\rm B} - \rho \tilde{\lambda}^{\rm B} \!\! & = & \!\!
     m^{\mu} m^{\nu} T_{\mu\nu}, \nonumber \\
\overline{\Psi}^{\rm B}_{4} - (\delta - 2 \beta) \tilde{\nu}^{\rm B} +
     (\Delta + 2 \gamma + 2 \mu) \tilde{\lambda}^{\rm B} \!\! & = & \!\! 0,
     \nonumber \\
- \tilde{\Psi}^{\rm B}_{3} + D \tilde{\nu}^{\rm B} - (\Delta + \mu)
     \tilde{\pi}^{\rm B} - \mu \tilde{\tau}^{\rm B} - \Delta \phi
     \tilde{\phi}^{\rm B} \!\! & = & \!\! n^{\mu} m^{\nu} T_{\mu\nu},
\end{eqnarray}
and linearizing Bianchi identities:
\begin{eqnarray}
\!\! & & \!\! (\overline{\delta} + 4 \overline{\beta}) \Psi^{\rm
     B}_{0} - (D - 4 \rho) \tilde{\Psi}^{\rm B}_{1} - (3 \Psi_{2} + 2
     D\phi \Delta\phi - 2 \Phi_{11})\tilde{\kappa}^{\rm B} \nonumber \\
\!\! & & \!\! + (D\phi)^{2} \tilde{\pi}^{\rm B} - (D\phi)^{2} D
     (\tilde{\phi}^{\rm B}/ D\phi) = - (D - 2 \rho) l^{\mu} m^{\nu}
     T_{\mu\nu} + \delta l^{\mu} l^{\nu} T_{\mu\nu},  \nonumber \\
\!\! & & \!\! (\overline{\delta} + 4 \overline{\beta})
     \overline{\Psi}^{\rm B}_{4} - (\Delta + 2 \gamma + 4 \mu)
     \tilde{\Psi}^{\rm B}_{3} + (3 \Psi_{2} - 2 \Phi_{11} + 2 D\phi
     \Delta\phi) \tilde{\nu}^{\rm B} - (\Delta\phi)^{2} \tilde{\tau}^{\rm B}
     \nonumber \\
\!\! & & \!\! -(D\phi \Delta\phi) \Delta(\tilde{\phi}^{\rm B}/ D\phi) =
     - (\Delta + 2 \gamma + 2 \mu) m^{\mu} n^{\nu} T_{\mu\nu} + \delta
     (n^{\mu} n^{\nu} T_{\mu\nu}).
\end{eqnarray}

Thus, we have finally a complete system of thirteen equations (A6), (A7),
(A12), (A14), (A20)-(A22) for thirteen unknowns, the nine ones given in
(A3), plus $\Psi^{\rm B}_{0}$, $\overline{\Psi}^{\rm B}_{4}$, 
$\tilde{\Psi}^{\rm B}_{1}$, and $\tilde{\Psi}^{\rm B}_{3}$. All the other 
equations appear to be a consequence of them. It is worth to point out that 
if one considers directly perturbation equations such as (A5), (A10), (A11), 
and (A19), without involving gauge invariance quantities, then, one obtains 
a system of equations in which the number of unknowns exceed highly the 
number of possible equations. Therefore, apparently there is a direct 
physical meaning behind the existence of the complete system obtained here; 
it is what may be obtained in a form that involves only certain natural 
gauge invariant perturbed field quantities. However, the system for 
thirteen unknowns, will be no used as obtained, but a more manageable 
system is obtained from it in Sec.\ III. For this purpose, the two 
following equations are useful, which come from the combinations of Eqs.\ 
(A21), or directly from linearizing Ricci identities:
\begin{eqnarray}
(\Delta - 4 \gamma + \mu) \Psi^{\rm B}_{0} - (\delta - 2 \beta)
     \tilde{\Psi}^{\rm B}_{1} - (3 \Psi_{2} + 2 \Phi_{11})
     \tilde{\sigma}^{\rm B} - D \phi (\delta - 2 \beta) \tilde{\phi}^{\rm B}
     + (D\phi)^{2} \tilde{\lambda}^{\rm B} \nonumber \\
     = (\delta - 2 \beta) l^{\mu} m^{\nu} T_{\mu\nu} - (D - \rho) m^{\mu}
     m^{\nu} T_{\mu\nu}, \nonumber \\
(D - \rho) \overline{\Psi}^{\rm B}_{4} - (\delta - 2 \beta)
     \tilde{\Psi}^{\rm B}_{3} + (3 \Psi_{2} + 2 \Phi_{11})
     \tilde{\lambda}^{\rm B} - \Delta\phi (\delta - 2 \beta)
     \tilde{\phi}^{\rm B} - (\Delta\phi)^{2} \tilde{\sigma}^{\rm B}
     \nonumber \\
     = (\delta - 2 \beta) n^{\mu} m^{\nu} T_{\mu\nu} - (\Delta + \mu)
     m^{\mu} m^{\nu} T_{\mu\nu}.
\end{eqnarray}
\\[2em]

\begin{center}
{\uno Appendix B: Separation of variables for the field variations}
\end{center}
\vspace{1em}
\renewcommand{\theequation}{B\arabic{equation}}
\setcounter{equation}{0}
\vspace{1em}

The separation of variables for the potentials in Eq.\ (51) implies a
separation for the components of the field variations. For example, from
Eqs.\ (39) and (40) (considering that the only nonvanishing contractions
of the tetrad $(l_{\mu}, n_{\mu}, m_{\mu}, \overline{m}_{\mu})$ are
$l^{\mu} n_{\mu} = 1 = - m^{\mu} \overline{m}_{\mu}$), $l^{\mu} b_{\mu} =
- \frac{1}{4\xi} [(\delta + 2 \beta)(\delta + 4 \beta)
\frac{\psi_{F}}{\varphi_{1}} + c.c.]$, which reduces to
\begin{equation}
    l^{\mu} b_{\mu} = \frac{i L[l(l+1)]^{1/2}}{4Q \xi} \left[ \psi_{f} \, 
    Y_{lm} \, e^{-i\omega t} - c.c. \right],
\end{equation}
where we have employed the second of Eqs.\ (7), Eq.\ (51), and repeatedly
the first of Eqs.\ (50). From Eq.\ (B1), and using again the first of
Eqs.\ (50) we obtain the following useful expression
\begin{equation}
    (\delta - 2 \beta) \delta (l^{\mu} b_{\mu}) = 
    \frac{i L^{2}l(l+1)}{8Q\xi R^{2}} \left[ \psi_{f} \, {_{2}}Y_{lm} \, 
    e^{-i\omega t} - \overline{\psi}_{f} \, \overline{{_{-2}}Y_{lm}} \,
    e^{i\omega t} \right] ,
\end{equation}
and similarly for the other components of the electromagnetic field
variations:
\begin{eqnarray}
n^{\mu} b_{\mu} \!\! & = \!\! & - \frac{1}{4 \xi} \left[ (\delta + 2
    \beta)(\delta + 4 \beta) \frac{\psi_{E}}{\varphi_{1}} + c.c. \right]  
    = \frac{iL[l(l+1)]^{1/2}}{4Q \xi}  \left[ \psi_{e} \, Y_{lm} \, 
    e^{-i\omega t} - c.c. \right], \nonumber \\
(\delta - 2 \beta) \delta (n^{\mu} b_{\mu}) \!\! & = & \!\! 
    \frac{iL^{2}l(l+1)}{8Q \xi R^{2}} \left[ \psi_{e} \, {_{2}}Y_{lm} 
    \, e^{-i\omega t} - \overline{\psi}_{e} \, \overline{{_{-2}}Y_{lm}} \,
    e^{i\omega t} \right] , \nonumber \\
m^{\mu} b_{\mu} \!\! & = & \!\! - \frac{1}{4 \xi} [(D + \rho)
    (\overline{\delta} + 4 \overline{\beta})
    \frac{\overline{\psi}_{E}}{\overline{\varphi}_{1}} + (\Delta - 2
    \gamma - \mu) (\overline{\delta} + 4 \overline{\beta})
    \frac{\overline{\psi}_{F}}{\overline{\varphi}_{1}} - 8 a
    \varphi_{1} \xi (\overline{\delta} + 4 \overline{\beta})
    \overline{\psi}_{D}], \nonumber \\
(\delta - 2 \beta) (m^{\mu} b_{\mu}) \!\! & = & \!\!
    \frac{iL^{2}}{4Q\xi} \left[ \overline{\cal D} \, \overline{\psi}_{e} 
    - (\frac{\chi^{2}}{2} {\cal D} + 2 \gamma) \overline{\psi}_{f} + 8 a
    \xi \varphi^{2}_{1} \, \overline{\psi}_{d} \right] 
    \overline{{_{-1}}Y_{lm}} \, e^{i\omega t}.
\end{eqnarray}
For the components of the metric variations, using Eqs.\ (39) and (40), we 
have the expressions:
\begin{eqnarray}
\frac{1}{2} l^{\mu} l^{\nu} h_{\mu\nu} \!\! & = & \!\! (\delta + 2
    \beta) (\delta + 4 \beta) \psi_{H} + c.c. = 
    \frac{L[l(l+1)]^{1/2}}{2R^{2}}  \left[ \psi_{h} \, Y_{lm} \,
    e^{-i\omega t} + c.c. \right], \nonumber \\
(\delta - 2 \beta) \delta (l^{\mu} l^{\nu} h_{\mu\nu}) \!\! & = & \!\!
    \frac{L^{2}l(l+1)}{2R^{4}}  \left[ \psi_{h} \, {_{2}}Y_{lm} \,
    e^{-i\omega t} + \overline{\psi}_{h} \,  \overline{{_{-2}}Y_{lm}} \,
    e^{i\omega t} \right], \nonumber \\
\frac{1}{2} n^{\mu} n^{\nu} h_{\mu\nu} \!\! & = & \!\! (\delta + 2
    \beta) (\delta + 4 \beta) \psi_{G} + c.c. =
    \frac{L[l(l+1)]^{1/2}}{2R^{2}} \left[ \psi_{g} \, Y_{lm} \,
    e^{-i\omega t} + c.c. \right], \nonumber \\
(\delta - 2 \beta) \delta (n^{\mu} n^{\nu} h_{\mu\nu}) \!\! & = & \!\! 
    \frac{L^{2}l(l+1)}{2R^{4}}  \left[ \psi_{g} \, {_{2}}Y_{lm} \,
    e^{-i\omega t} + \overline{\psi}_{g} \,  \overline{{_{-2}}Y_{lm}} \,  
    e^{i\omega t} \right], \nonumber \\
\frac{1}{2} l^{\mu} m^{\nu} h_{\mu\nu} \!\! & = & \!\! (\Delta  - 4
   \gamma - \mu) (\overline{\delta} + 4 \overline{\beta})
    \overline{\psi}_{\rm H} - (\overline{\delta} + 4 \overline{\beta})
    \overline{\psi}_{\rm F} - D \phi (\overline{\delta} + 4
    \overline{\beta}) \overline{\psi}_{\rm D} \nonumber \\
    \!\! & = & \!\! - \frac{L}{\sqrt{2}} \left[ (\frac{\chi^{2}}{2} {\cal
    D} + 4 \gamma + \mu) \frac{\overline{\psi}_{h}}{R} + 
    \frac{\overline{\psi}_{f}}{R} + D \phi \frac{\overline{\psi}_{d}}{R} 
    \right] \overline{{_{-1}}Y_{lm}} \,  e^{i\omega t},  \nonumber \\
(\delta - 2 \beta) (l^{\mu} m^{\nu} h_{\mu\nu}) \!\! & = & \!\! 
    \frac{L^{2}}{R^{2}} \left[ ( \frac{\chi^{2}}{2} {\cal D} + 4
    \gamma + 2 \mu) \overline{\psi}_{h} + \overline{\psi}_{f} + D \phi
    \overline{\psi}_{d} \right] \overline{{_{-2}}Y_{lm}} \,
    e^{i\omega t}, \nonumber \\
\frac{1}{2} n^{\mu} m^{\nu} h_{\mu\nu} \!\! & = & \!\! (D + \rho)
    (\overline{\delta} + 4 \overline{\beta}) \overline{\psi}_{\rm G} -
    (\overline{\delta} + 4 \overline{\beta}) \overline{\psi}_{\rm E}
    - \Delta\phi (\overline{\delta} + 4 \overline{\beta})
    \overline{\psi}_{\rm D} \nonumber \\
    \!\! & = & \!\! \frac{L}{\sqrt{2}} \left[ (\overline{{\cal D}} + \rho)
    \frac{\overline{\psi}_{g}}{R} - \frac{\overline{\psi}_{e}}{R}  
    - \Delta\phi \frac{\overline{\psi}_{d}}{R} \right] 
    \overline{{_{-1}}Y_{lm}} \, e^{i\omega t},  \nonumber \\
(\delta - 2 \beta) (n^{\mu} m^{\nu} h_{\mu\nu}) \!\! & = & \!\! -
    \frac{L^{2}}{R^{2}} [(\overline{{\cal D}} + 2 \rho)
    \overline{\psi}_{g} - \overline{\psi}_{e} - \Delta\phi
    \overline{\psi}_{d}] \overline{{_{-2}}Y_{lm}} \, e^{i\omega t}, 
    \nonumber \\
\frac{1}{2} m^{\mu} m^{\nu} h_{\mu\nu} \!\! & = & \!\!  [(D - \rho) (D
    + 3 \rho) + \Phi_{00}] \overline{\psi}_{\rm G} + [(\Delta - 2 \gamma + 
    \mu) (\Delta - 4 \gamma - 3 \mu) + \Phi_{22}] \overline{\psi}_{\rm H} 
    \nonumber \\
    \!\! & & \!\! - 2 (D - \rho) \overline{\psi}_{\rm E} - 2 (\Delta -
    2 \gamma + \mu) \overline{\psi}_{\rm F} - [8 a \xi \varphi^{2}_{1}
    + \Delta\phi D + D\phi \Delta] \overline{\psi}_{\rm D}
    \nonumber \\
    \!\! & = & \!\! \left\{ [(\overline{{\cal D}} - \rho)
    (\overline{{\cal D}} + 3 \rho) + \Phi_{00}] \overline{\psi}_{g}
    + [(-\frac{\chi^{2}}{2} {\cal D} - 2 \gamma + \mu)
    (- \frac{\chi^{2}}{2} {\cal D} - 4 \gamma - 3 \mu) \right. 
    \nonumber \\
    \!\! & & \!\! + \left. \Phi_{22}] \overline{\psi}_{h} - 2
    (\overline{{\cal D}} - \rho) \overline{\psi}_{\rm e} - 2 (-
    \frac{\chi^{2}}{2} {\cal D} - 2 \gamma + \mu) \overline{\psi}_{f}
    \right. \nonumber \\
    \!\! & & \!\! - \left. [8 a \xi \varphi^{2}_{1} + \Delta \phi
    \overline{{\cal D}} - \frac{\chi^{2}}{2} D\phi {\cal D}]
    \overline{\psi}_{d} \right\}  \overline{{_{-2}}Y_{lm}} \, e^{i\omega
    t}, \\
\noalign{\hbox{and}}
    \phi^{\rm B} \!\! & = & \!\! \frac{1}{2} (\delta + 2 \beta) (\delta +
    4 \beta) \psi_{\rm D} + c.c. = \frac{L[l(l+1)]^{1/2}}{4R^{2}} [ 
    \psi_{d} Y_{lm} \, e^{-i\omega t} + c.c.], \nonumber \\
(\delta - 2 \beta) \delta \phi^{\rm B} \!\! & = & \!\!
    \frac{L^{2}l(l+1)}{8R^{4}}  \left[ \psi_{d} \, {_{2}}Y_{lm} \,
    e^{-i\omega t} + \overline{\psi}_{d} \,  \overline{{_{-2}}Y_{lm}} \,  
    e^{i\omega t}  \right], 
\end{eqnarray}
for dilaton field variations.

As we have seen, all gauge invariant variations of the Newman-Penrose
quantities are defined in terms of the components of the field variations
given in Eqs.\ (B1)--(B5). Particularly, from Eqs. (A1)--(A3) we have that
\begin{eqnarray}
\tilde{\nu}^{\rm B} \!\! & = & \!\! -(\Delta + 2 \gamma + \mu) (m^{\nu}
    n^{\nu} h_{\mu\nu}) + \frac{1}{2} \delta (n^{\mu} n^{\nu} h_{\mu\nu})
    + \frac{1}{2} (\Delta + 2 \gamma + \mu) \frac{1}{\varphi_{1}} [\delta
    (n^{\mu} b_{\mu}) - (\Delta + \mu) (m^{\mu}b_{\mu})], \nonumber \\
\tilde{\kappa}^{\rm B} \!\! & = & \!\! (D - \rho) (l^{\mu} m^{\nu}
    h_{\mu\nu}) - \frac{1}{2} \delta (l^{\mu} l^{\nu} h_{\mu\nu}) +
    \frac{1}{2} (D - \rho) \frac{1}{\varphi_{1}} [ (D - \rho) (m^{\mu}
    b_{\mu}) - \delta (l^{\mu} b_{\mu})], \nonumber \\
\tilde{\sigma}^{\rm B} \!\! & = & \!\! D \left( \frac{1}{2} m^{\mu} m^{\nu}
    h_{\mu\nu} \right) + \frac{1}{2\varphi_{1}} (\delta - 2 \beta) [(D -
    \rho) (m^{\mu} b_{\mu}) - \delta (l^{\mu} b_{\mu})], \nonumber \\
\tilde{\lambda}^{\rm B} \!\! & = & \!\! - \Delta \left( \frac{1}{2}
    m^{\mu} m^{\nu} h_{\mu\nu} \right) + \frac{1}{2\varphi_{1}} (\delta -
    2 \beta) [\delta (n^{\mu} b_{\mu}) - (\Delta + \mu) (m^{\mu} b_{\mu})],
\end{eqnarray}
and from Eqs.\ (A21)
\begin{eqnarray}
\Psi^{\rm B}_{0} \!\! & = & \!\! - (\delta - 2 \beta) \tilde{\kappa}^{\rm B} 
    + (D - 2 \rho) \tilde{\sigma}^{\rm B} = (D - 2 \rho) D \left( 
    \frac{1}{2} m^{\mu} m^{\nu} h_{\mu\nu} \right) - (D - 2 \rho) [(\delta 
    - 2 \beta) (l^{\mu}m^{\nu} h_{\mu\nu})] \nonumber \\
    \!\! & & \!\! + \frac{1}{2} (\delta - 2 \beta) \delta (l^{\mu} l^{\nu}
    h_{\mu\nu}), \nonumber \\
\overline{\Psi}^{\rm B}_{4} \!\! & = & \!\! (\delta - 2 \beta)
    \tilde{\nu}^{\rm B} - (\Delta + 2 \gamma + 2 \mu)
    \tilde{\lambda}^{\rm B} = (\Delta + 2 \gamma + 2 \mu) \Delta \left(
    \frac{1}{2} m^{\mu} m^{\nu} h_{\mu\nu} \right) \nonumber \\
    \!\! & & \!\! - (\Delta + 2 \gamma + 2 \mu)
    [(\delta - 2 \beta) (n^{\mu} m^{\nu} h_{\mu\nu})] + \frac{1}{2}
    (\delta - 2 \beta) \delta (n^{\mu} n^{\nu} h_{\mu\nu}).
\end{eqnarray}

Hence, substituting directly Eqs.\ (B1)--(B4) into Eqs.\ (B6) and (B7), we
have the following expressions for the quantities appearing in the
decoupled system:
\begin{eqnarray}
\Psi^{\rm B}_{0} \!\! & = & \!\! \left[ (\overline{{\cal D}} - 2 \rho) 
    \overline{{\cal D}} \left( \frac{1}{2} m^{\mu} m^{\nu} h_{\mu\nu} 
    \right) (r) - (\overline{{\cal D}} - 2 \rho) [(\delta - 2 \beta) 
    (l^{\mu} m^{\nu} h_{\mu\nu})] (r) + \frac{L^{2}l(l+1)}{4R^{4}} 
    \overline{\psi}_{h} \right] \overline{{_{-2}}Y_{lm}} \, e^{i\omega t} 
    \nonumber \\
    \!\! & & \!\! + \frac{L^{2}l(l+1)}{4R^{4}} \psi_{h} \, {_{2}}Y_{lm}
    \, e^{-i\omega t} \equiv  \Psi^{{\rm B}+}_{0} \overline{{_{-2}}Y_{lm}} 
    \, e^{i\omega t} + \Psi^{{\rm B}-}_{0} \, {_{2}}Y_{lm} \, e^{-i\omega 
    t}, \nonumber \\
\overline{\Psi}^{\rm B}_{4} \!\! & = & \!\! \left\{ - \frac{1}{2}
    ( - \frac{\chi^{2}}{2} {\cal D} + 2 \gamma + 2 \mu) \chi^{2} {\cal D} 
    (\frac{1}{2} m^{\mu} m^{\nu} h_{\mu\nu}) (r) - ( - \frac{\chi^{2}}{2} 
    {\cal D} + 2 \gamma + 2 \mu) [(\delta - 2 \beta) (n^{\mu} m^{\nu} 
    h_{\mu\nu})] (r) \right. \nonumber \\
    \!\! & & \!\! \left. + \frac{L^{2}l(l+1)}{4R^{4}}
    \overline{\psi}_{g} \right\} \overline{{_{-2}}Y_{lm}} \,
    e^{i\omega t} + \frac{L^{2}l(l+1)}{4R^{4}} \psi_{g} \, {_{2}}Y_{lm} \, 
    e^{-i\omega t} \equiv \overline{\Psi}^{{\rm B}+}_{4} \, 
    \overline{{_{-2}}Y_{lm}} \, e^{i\omega t} + \overline{\Psi}^{{\rm
    B}-}_{4} \, {_{2}}Y_{lm} \, e^{-i\omega t}, \nonumber \\
\tilde{\sigma}^{\rm B} \!\! & = & \!\! \left\{ \overline{\cal D}
    (\frac{1}{2} m^{\mu} m^{\nu} h_{\mu\nu}) (r) + \frac{1}{2
    \varphi_{1}} (\overline{\cal D} - 2 \rho) [(\delta - 2 \beta)
    m^{\mu} b_{\mu}] (r) + \frac{L^{2}l(l+1)}{8 Q^{2} \xi}
    \overline{\psi}_{f} \right\} \overline{{_{-2}}Y_{lm}} \,
    e^{i\omega t} \nonumber \\
    \!\! & & \!\! - \frac{L^{2}l(l+1)}{8 Q^{2} \xi} \psi_{f} \, 
    {_{2}}Y_{lm} \, e^{-i\omega t} \equiv \tilde{\sigma}^{{\rm B}+} (r) 
    \overline{{_{-2}}Y_{lm}} \, e^{i\omega t} + \tilde{\sigma}^{{\rm B}-} 
    (r) \, {_{2}}Y_{lm} \, e^{-i\omega t}, \nonumber \\
\tilde{\lambda}^{\rm B} \!\! & = & \!\! \left\{ \frac{\chi^{2}}{2} {\cal
    D} ( \frac{1}{2} m^{\mu} m^{\nu} h_{\mu\nu}) (r) -
    \frac{1}{2\varphi_{1}} ( - \frac{\chi^{2}}{2} {\cal D} + 2 \mu) 
    ((\delta - 2 \beta) m^{\mu} b_{\mu}) (r) - \frac{L^{2}l(l+1)}{8 Q^{2} 
    \xi} \overline{\psi}_{e} \right\} \overline{{_{-2}}Y_{lm}} \, 
    e^{i\omega t} \nonumber \\
    \!\! & & \!\! + \frac{L^{2}l(l+1)}{8 Q^{2} \xi} \psi_{e} \,
    {_{2}}Y_{lm} \, e^{-i\omega t} \equiv \tilde{\lambda}^{{\rm B}+} (r) 
    \overline{{_{-2}}Y_{lm}} \, e^{i\omega t} + \tilde{\lambda}^{{\rm B}-} 
    (r) \, {_{2}}Y_{lm} \, e^{-i\omega t},
\end{eqnarray}
where $(r)$ denotes the radial part of the corresponding quantity. For
example, from Eqs.\ (B4), $[(\delta - 2 \beta) (n^{\mu}m^{\nu}
h_{\mu\nu})] (r) = - \frac{L^{2}}{R^{2}} [(\overline{{\cal D}} + 2 \rho) 
\overline{\psi}_{g} - \overline{\psi}_{e} - \Delta\phi 
\overline{\psi}_{d}]$, and similarly for $[(\delta - 2 \beta) (l^{\mu} 
m^{\nu} h_{\mu\nu})] (r)$, $(\frac{1}{2} m^{\mu} m^{\nu} h_{\mu\nu}) 
(r)$, and $[(\delta - 2 \beta) m^{\mu} b_{\mu}] (r)$ from Eqs.\ (B4) and 
(B3). Moreover, the second equalities are only for defining in a compact 
way the radial parts of the form $e^{i\omega t}$ and $e^{-i\omega t}$ of 
the corresponding quantity. Finally, from the last of Eqs.\ (A3) and (B5) 
we have that
\begin{eqnarray}
(\delta - 2 \beta) \tilde{\phi}^{\rm B} \!\! & = & \!\! (\delta - 2
    \beta) \delta\phi^{\rm B} -  \frac{1}{2\varphi_{1}} [\Delta\phi (D -
    2 \rho) + D\phi (\Delta + 2 \mu)] (\delta - 2\beta) (m^{\mu} b_{\mu})
    \nonumber \\
    \!\! & & \!\! + \frac{1}{2\varphi_{1}} (\delta - 2 \beta) \delta [ 
    \Delta\phi (l^{\mu} b_{\mu}) + D\phi (n^{\mu} b_{\mu})] \nonumber \\
\!\! & = & \!\! \left\{ - \frac{1}{2\varphi_{1}} [\Delta\phi 
    (\overline{{\cal D}} - 2 \rho) + D\phi (- \frac{\chi^{2}}{2} {\cal D} 
    + 2\mu)] [(\delta - 2\beta) m^{\mu} b_{\mu} (r)] \right. \nonumber \\
    \!\! & & \!\! \left. - \frac{L^{2}l(l+1)}{8} \left[ -
    \frac{\overline{\psi}_{d}}{R^{4}} + \frac{1}{Q^{2}\xi} (D\phi
    \overline{\psi}_{e} + \Delta\phi \overline{\psi}_{f}) \right]
    \right\} \overline{{_{-2}}Y_{lm}} \, e^{i\omega t} \nonumber \\
    \!\! & & \!\! +  \frac{L^{2}l(l+1)}{8} \left\{ \frac{\psi_{d}}{R^{4}} 
    + \frac{1}{Q^{2}\xi} (D\phi \psi_{e} + \Delta\phi \psi_{f}) \right\}
    {_{2}}Y_{lm} \, e^{-i \omega t}  \nonumber \\
    \!\! & \equiv & \!\! \tilde{\phi}^{{\rm B}+} \, 
    \overline{{_{-2}}Y_{lm}} \, e^{i\omega t} + \tilde{\phi}^{{\rm B}-} \,
    {_{2}}Y_{lm} \, e^{-i\omega t}.
\end{eqnarray}

From Eqs.\ (B8) and (B9) we can obtain the following useful relations:
\begin{eqnarray}
    \psi_{g} \Psi^{{\rm B}-}_{0} - \psi_{h} \overline{\Psi}^{{\rm B}-}_{4}
    \!\! & = & \!\! 0, \nonumber \\
    \psi_{e} \tilde{\sigma}^{{\rm B}-} + \psi_{f}\tilde{\lambda}^{{\rm
    B}-} \!\! & = & \!\! 0, \nonumber \\
    \Delta\phi \psi_{d} \tilde{\sigma}^{{\rm B}-} - D\phi \psi_{d}
    \tilde{\lambda}^{{\rm B}-} + \psi_{d} \tilde{\phi}^{{\rm B}-}
    \!\! & = & \!\! \frac{L^{2}l(l+1)}{8R^{4}} \psi^{2}_{d}.
\end{eqnarray}
Note that $(\Psi^{\rm B})^{+}$ depends on $(\overline{\psi}_{i})$, whereas
$(\Psi^{\rm B})^{-}$ on $(\psi_{i})$. \\[2em]

}
\end{document}